\documentclass[preprint,prd,showpacs,preprintnumbers,amsmath,amssymb,aps,floatfix,nofootinbib]{revtex4-1}
\usepackage[colorlinks=true,linkcolor=blue,filecolor=blue,urlcolor=blue,citecolor=blue,pdftex=true,plainpages=false]{hyperref}
\usepackage{amsmath,amssymb,amsbsy,amsfonts,bm}
\usepackage{graphicx}
\usepackage[mathscr]{euscript}
\usepackage{multirow}

\usepackage{color}
\usepackage{bm}

\usepackage{graphicx}
\usepackage{subfigure}

\def\chpt{\raise0.4ex\hbox{$\chi$}PT}
\def\pqchpt{PQ\raise0.4ex\hbox{$\chi$}PT}
\def\schpt{S\raise0.4ex\hbox{$\chi$}PT}
\def\rschpt{rS\raise0.4ex\hbox{$\chi$}PT}
\def\cpt{\raise0.4ex\hbox{$\chi$}PT}
\def\scpt{S\raise0.4ex\hbox{$\chi$}PT}
\def\rscpt{rS\raise0.4ex\hbox{$\chi$}PT}

\def\secref#1{Sec.~\ref{sec:#1}}
\def\secrefs#1#2{Secs.~\ref{sec:#1} and \ref{sec:#2}}

\def\gtwid{{\,\raise.3ex\hbox{$>$\kern-.75em\lower1ex\hbox{$\sim$}}\,}}
\def\ltwid{{\,\raise.3ex\hbox{$<$\kern-.75em\lower1ex\hbox{$\sim$}}\,}}

\def\ie{{\it i.e.},\ }
\def\eg{{\it e.g.},\ }

\def\cA{{\cal A}}

\def\cL{{\cal L}}
\def\cM{{\cal M}}

\def\cO{{\cal O}}

\def\cV{{\cal V}}

\def\rcite#1{Ref.~\cite{#1}}
\def\Rcite#1{Reference~\cite{#1}}
\def\rcites#1{Refs.~\cite{#1}}

\def\eqn#1{\label{eq:#1}}
\def\Equation#1{Equation~(\ref{eq:#1})}
\def\Equations#1#2{Equations~(\ref{eq:#1}) and (\ref{eq:#2})}
\def\EQuations#1#2#3{Equations~(\ref{eq:#1}), (\ref{eq:#2}) and (\ref{eq:#3})}
\def\eq#1{Eq.~(\ref{eq:#1})}
\def\eqsthru#1#2{Eqs.~(\ref{eq:#1}) through (\ref{eq:#2})}
\def\eqs#1#2{Eqs.~(\ref{eq:#1}) and (\ref{eq:#2})}
\def\eqsthree#1#2#3{Eqs.~(\ref{eq:#1}), (\ref{eq:#2}) and (\ref{eq:#3})}
\def\eqsfour#1#2#3#4{Eqs.~(\ref{eq:#1}), (\ref{eq:#2}), (\ref{eq:#3}) and (\ref{eq:#4})}

\newcommand{\diag}{\ensuremath{\operatorname{diag}}}

\newcommand*{\tr}{\ensuremath{\operatorname{tr}}}
\newcommand{\sTr}{\ensuremath{\textrm{s\kern-.09em Tr}}}
\def\str{{\rm str}}

\newcommand{\trD}{\ensuremath{\textrm{tr}_{\textrm{\tiny \it D}}}}
\renewcommand{\Re}{{\rm Re}}

\newcommand{\vslash}{\makebox[0pt][l]{/}v}

\def\gtwid{{\,\raise.3ex\hbox{$>$\kern-.75em\lower1ex\hbox{$\sim$}}\,}}
\def\ltwid{{\,\raise.3ex\hbox{$<$\kern-.75em\lower1ex\hbox{$\sim$}}\,}}

\newcommand{\bi}{\begin{itemize}}
\newcommand{\ei}{\end{itemize}}

\newcommand{\be}{\begin{equation}}
\newcommand{\ee}{\end{equation}}

\newcommand{\bea}{\begin{eqnarray}}
\newcommand{\eea}{\end{eqnarray}}

\newcommand{\bt}[1]{\begin{table}[!t] \begin{center}\begin{tabular}{#1} \hline\hline  \\[-0.5em]}
\newcommand{\et}[2]{\hline\hline \end{tabular} \end{center} \caption{#1} \label{#2} \end{table}}

\def\LP{\left(}         
\def\RP{\right)}        

\def\PARTWO#1#2{ {{\partial^2 #1}\over{\partial #2}^2} }

\newcommand{\BE}{\begin{displaymath}}
\newcommand{\EE}{\end{displaymath}}
\def\BNE{\begin{equation}}
\def\ENE{\end{equation}}
\def\BEA{\begin{eqnarray}}
\def\EEA{\nonumber\end{eqnarray}}
\def\EEAN{\end{eqnarray}}

\def\aschpt{HMrAS\raise0.4ex\hbox{$\chi$}PT}
\def\negcdot{\negmedspace\cdot\negmedspace}
\def\leftvec{{\raise1.5ex\hbox{$\leftarrow$}\kern-1.00em}}
\def\rightvec{{\raise1.5ex\hbox{$\rightarrow$}\kern-1.00em}}

\begin{document}

\title{Effects of non-equilibrated topological charge distributions on pseudoscalar meson masses and decay constants\\
 }

\author{C.~Bernard}
\email{cb@wustl.edu}
\affiliation{ Department of Physics, Washington University, St. Louis, MO 63130, USA}
\author{D.~Toussaint}
\email{doug@physics.arizona.edu}
\affiliation{ Physics Department, University of Arizona Tucson, AZ 85721, USA}

\collaboration{MILC Collaboration}
\date{\today}

\begin{abstract}

We study the effects of failure to 
equilibrate the squared topological charge $Q^2$ on lattice calculations of pseudoscalar
masses and decay constants.  The analysis is based on
chiral perturbation theory  calculations of the dependence of these
quantities on the QCD vacuum angle $\theta$.  For the light-light partially quenched case, we rederive
 the known chiral perturbation theory results of Aoki and Fukaya, but using the
nonperturbatively-valid chiral theory worked out by Golterman, Sharpe and
Singleton, and by Sharpe and Shoresh.
We then extend these calculations to heavy-light mesons.  Results when staggered taste-violations are
important are also presented.   
The derived $Q^2$ dependence is compared to that  of simulations  using the MILC collaboration's ensembles of lattices with four
flavors of HISQ dynamical quarks. We find agreement, albeit with large
statistical errors.   These results can be used to correct for the leading
effects of unequilibrated $Q^2$, or to make estimates of the systematic error
coming from the failure to equilibrate $Q^2$.
In an appendix, we  show that the partially
quenched chiral theory may be extended beyond a lower bound on valence masses
discovered by Sharpe and Shoresh.  Subtleties occurring when a sea-quark mass vanishes
are discussed in another appendix.
\end{abstract}
\maketitle

\section{Introduction} \label{sec:introduction}

In continuum QCD the topological charge $Q$ cannot change in a continuous evolution
of the gluon fields.  Thus we expect that lattice QCD simulations using approximately
continuous evolution algorithms should see very slow evolution of the topological
charge, since changing the topological charge involves a tunneling where some of
the plaquettes or other loops in the gauge action pass through large values.
This expected slow evolution of the topological charge has been observed and studied
in Refs~\cite{DelDebbio:2001kz,Luscher:2010iy,MILCtopology2010,Schaeffer2011,McGlynn2014}.
Since the rate at which the topological charge $Q$ changes in a lattice simulation
falls off quickly as the lattice spacing decreases, modern QCD simulations are
reaching a regime where the distribution of $Q$ cannot be accurately sampled
in a simulation with practical length.
When this is the case, physical quantities will suffer a systematic error, and
we need to either correct for this error or account for it in our error budgets.

Here, we use the MILC collaboration's ensembles of lattices
with a one-loop Symanzik and tadpole improved gauge action and four flavors
of highly improved staggered quarks (HISQ) to study the errors induced by an insufficiently sampled
$Q$ distribution.  We first  demonstrate the expected slow evolution
of topological charge as the lattice spacing decreases.

We proceed to  calculate, in chiral perturbation theory (\chpt), the dependence of the
light-light and heavy-light pseudoscalar masses and decay constants
on the QCD vacuum angle $\theta$, which is related to their dependence on the
average $Q^2$ in the lattice simulation \cite{brower2003}.  The unitary case (valence and sea quark masses identical) for light-light mesons is treated first, mainly to introduce the methods and set the notation; the results 
already appear in \rcite{brower2003}, or may be obtained by straightforward generalization of that
calculation.  We then discuss light-light partially quenched case, which has also been
treated by Aoki and Fukaya \cite{aokifukaya2009} using   partially quenched chiral perturbation theory (\pqchpt)
with the the replica method \cite{REPLICA}.
 Because the vacuum state changes in
the presence of $\theta$, it could in principle be important to use a nonperturbatively valid method
for the partially quenched theory.  Rather than
the replica method \cite{REPLICA}, which has not been justified nonperturbatively,
 we therefore employ the approach 
to \pqchpt\ introduced by 
Golterman, Sharpe, and Singleton \cite{gss} and Sharpe and Shoresh \cite{sharpeshoresh2001}.
A potential sticking point, however, is the bound on the values of valence and sea quark masses
found by Sharpe and Shoresh.  When this bound is violated, the \pqchpt\ approach of 
\rcites{gss,sharpeshoresh2001} appears to break down.  We are able to show (Appendix \ref{ghost-mass})
that the bound is actually spurious, and the chiral theory continues to be valid when the bound is violated.
Once the partially quenched light-light case is analyzed, it is not difficult to generalize it to the
 heavy-light case, or to include the leading discretization effects coming from staggered
 taste violations.  The details of the partially quenched light-light and heavy-light calculations
constitute the majority of this paper.

Once the \chpt\ predictions are in hand, they are  compared to the HISQ simulation data.  
Although the statistical errors are large, we find good qualitative agreement between predictions and data.  We discuss how to use this information
to correct for the difference between the average of the squared topological charge in a simulation, $\langle Q^2\rangle_{sample}$,
 and the correct average $\langle Q^2\rangle$.

The remainder of this paper is organized as follows.
In section~\ref{sec:evolution}, we discuss the evolution of topological charge in our simulations.
 The connection between the dependence of physical quantities on the topological charge in fixed volume,
 and their dependence on $\theta$ in 
infinite volume, is reviewed in section~\ref{sec:fixedQ}.
Sections~\ref{sec:unitary} and~\ref{sec:PQ} present the dependence of the mass and decay constants of light-quark
pseudoscalar mesons in the unitary and partially
quenched cases, respectively.
In section~\ref{sec:heavy-light}, the calculation is extended to heavy-light pseudoscalar mesons.
We briefly describe the inclusion of staggered taste-violating effects in \secref{staggered}, and give the results
for the light-light and heavy-light cases.
Finally, in section~\ref{sec:compare}, we use the correlation between $Q^2$ and masses and decay
constants in our simulations to estimate the derivatives with respect to $\theta$, compare
them to \chpt, and discuss how simulation results might be adjusted.  A brief conclusion summarizes
our main results.

There are three appendices:  Appendix \ref{m=0} discusses subtleties that occur when one or more
sea-quark mass vanishes.  Appendix \ref{ghost-mass} investigates the Sharpe-Shoresh 
bound  \cite{sharpeshoresh2001} 
on quark masses in the partially quenched chiral theory.  A brief discussion of decoupling
issues in the current context  is presented in Appendix \ref{decoupling}.

A preliminary report of this work, 
which did not yet include results for heavy-light mesons or taste violations, was presented at {\it Lattice 2016} \cite{cbdtlat16}.

\section{Evolution of the topological charge in lattice simulations} \label{sec:evolution}

The ensembles we study have lattice spacings ranging from $0.09$ fm to $0.03$ fm,
and light sea quark masses ($m_l$) at either one fifth of the strange quark mass ($m_s$) or
approximately the physical light quark mass, which is approximately $m_s/27$.  
See \rcites{Bazavov:2012xda,Bazavov:2017lyh} for
the parameters of the ensembles and the details of their generation.
We measured the topological charge on these ensembles using the procedure described
in Ref.~\cite{asq_topology}.  This procedure consists of three HYP smearings of the lattice \cite{hyp_smear}
followed by an integration of the correlator of an improved topological density operator \cite{topo_operator}.
In addition to the tests described in Ref.~\cite{asq_topology}, there is a recent study
comparing many methods of measuring the topological charge, finding generally good
consistency among the methods~\cite{topo_compare}.

\begin{figure}
\vspace{-1.30in} \hspace{-0.25in}
\includegraphics[width=1.05\textwidth]{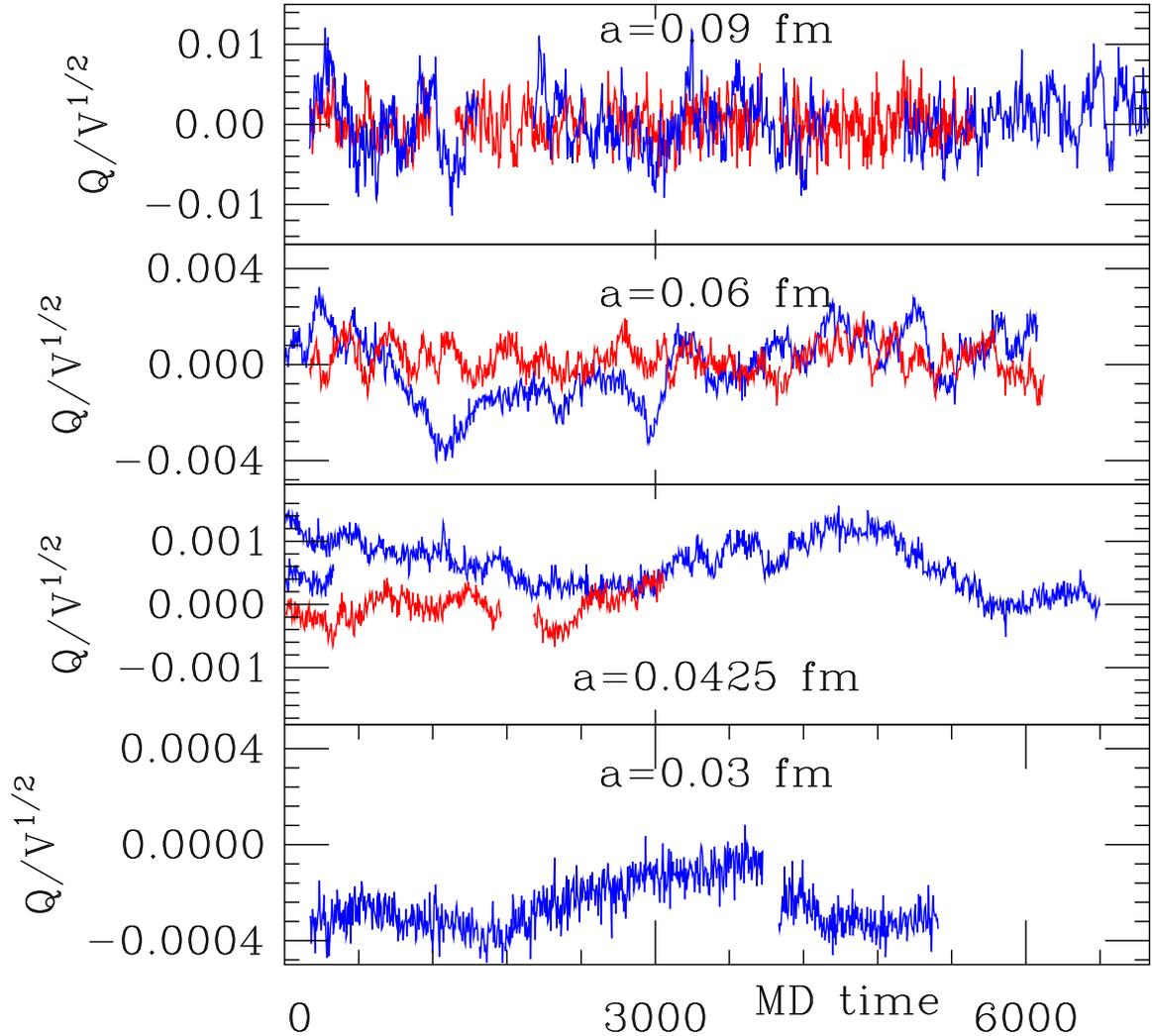}
\vspace{-2.20in}
\caption{
\label{topo_history_fig}
Topological charge time histories for various lattice spacings; note that the vertical scale decreases as the lattice spacing decreases (top to bottom). Blue lines are for ensembles
with light sea quark mass one fifth of the strange quark mass, and the red lines are for ensembles with
light sea quark mass at its physical value, $\approx\! m_s/27$.
Notice the narrower distributions and shorter autocorrelation times for physical quark mass ensembles.
Breaks in the traces separate multiple runs at the same couplings.
The second short blue trace at $a=0.0425$ fm 
is from a run with three times longer molecular dynamics trajectories than the main run.
}
\end{figure}

\begin{figure}
\vspace{-1.30in} \hspace{-0.25in}
\includegraphics[width=0.80\textwidth]{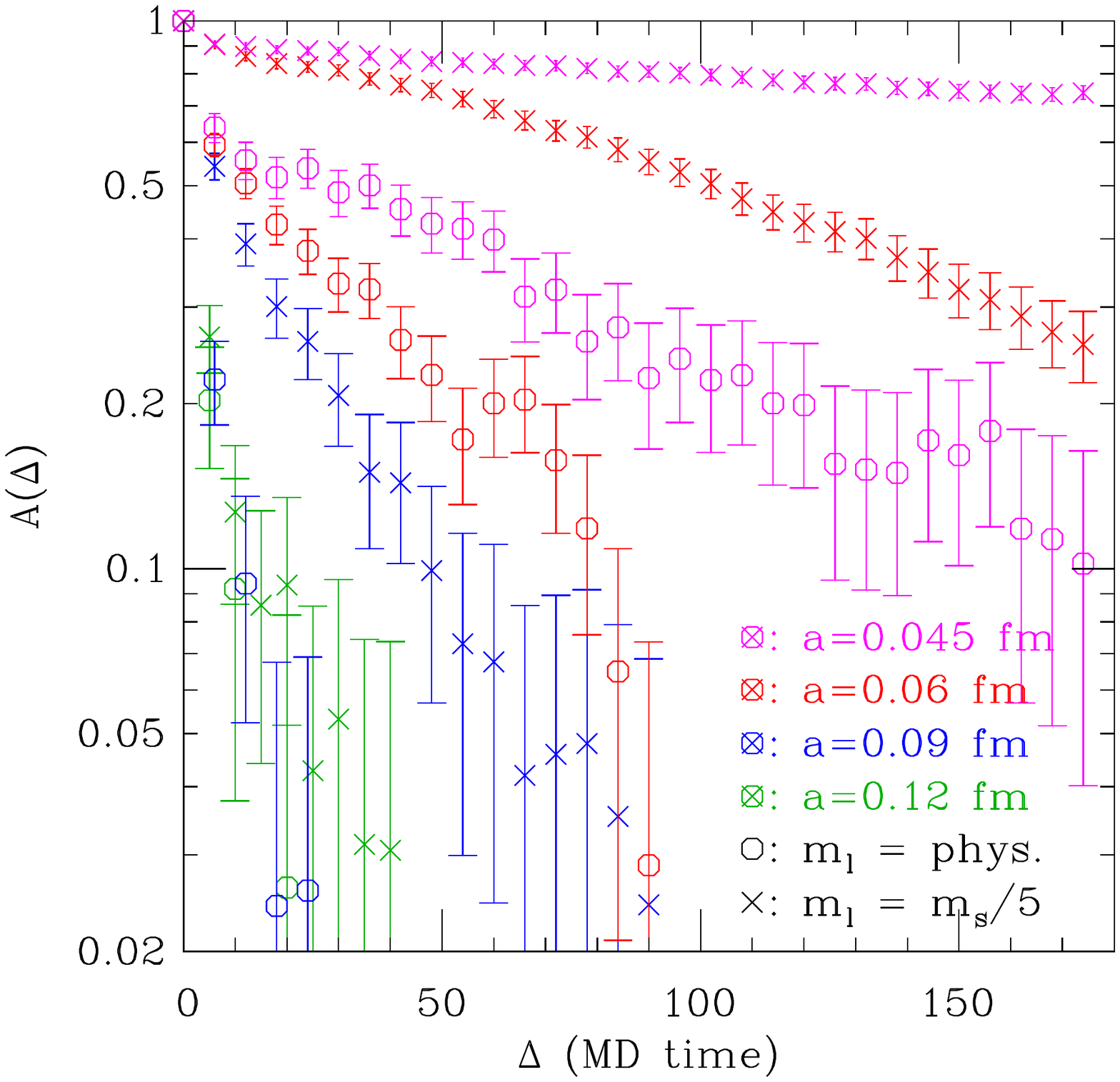}
\vspace{-1.30in}
\caption{
\label{fig:autocor}
Autocorrelations of the squared topological charge, where $A(\Delta)$ is defined in Eq.~\protect\ref{eq:Adef}.
For each lattice spacing, the crosses are the ensemble with $m_l=m_s/5$ and the octagons the ensemble with
$m_l$ at its physical value, $\approx\! m_s/27$.
}
\end{figure}

Figure~\ref{topo_history_fig} shows the time histories of $Q/V^{1/2}$ in our simulations, where
$V$ is the lattice volume in fm$\null^4$, and there are periodic boundary conditions on the 
gauge field in
all four directions.  In this plot the blue lines are for ensembles
with light sea quark mass one fifth of the strange quark mass, and the red lines are for ensembles with
physical light quark mass.  The increasing autocorrelation time of $Q$ as $a$
decreases is obvious, and at $a=0.03$ fm we see that the simulation has only
covered a small range of $Q^2$.
The operator we use to measure $Q$ is noisy enough, and the volume of the lattices large enough,
that we do not see plateaus at integer values of $Q$
in Fig.~\ref{topo_history_fig}, or, for that matter, in histograms of the topological charge.

For each lattice spacing, the local structure of the time evolution is similar
for the $m_l=m_s/5$ ensemble and the physical $m_l$ ensemble --- $Q$ typically changes by
about the same amount in each time unit.  However, in the
$m_l=m_s/5$ ensembles $Q$ ranges over larger values, and so it takes longer to random walk
over this range, leading to a longer autocorrelation time.  This is as
expected, since the gauge action controls the tunneling rate for $Q$, so
the average squared change in $Q$ per unit volume per unit simulation time is approximately
independent of the light quark mass.   However, the fermion determinant suppresses the
average $Q^2$, and the topological susceptibility, $\langle Q^2/V \rangle$,
is approximately proportional to $m_l$.
Figure~\ref{fig:autocor} shows the autocorrelation of the squared topological charge for four
different lattice spacings,
\BNE A(\Delta) = \frac{ \langle Q^2(t)Q^2(t+\Delta) \rangle - \langle Q^2(t) \rangle^2 }{  \langle Q^4(t) \rangle } .\label{eq:Adef} \ENE
We use $Q^2$ rather than $Q$ because it is $Q^2$ that controls
the effects on masses and decay constants (and all other CP conserving correlators).
In this graph we see the expected increase of the autocorrelation times as the
lattice spacing decreases, and also that the autocorrelations are smaller for the
physical quark mass ensembles (octagons) than for the $m_l=m_s/5$ ensembles (crosses).

\begin{figure}
\vspace{-1.35in} 
\begin{center}\includegraphics[width=.70\textwidth]{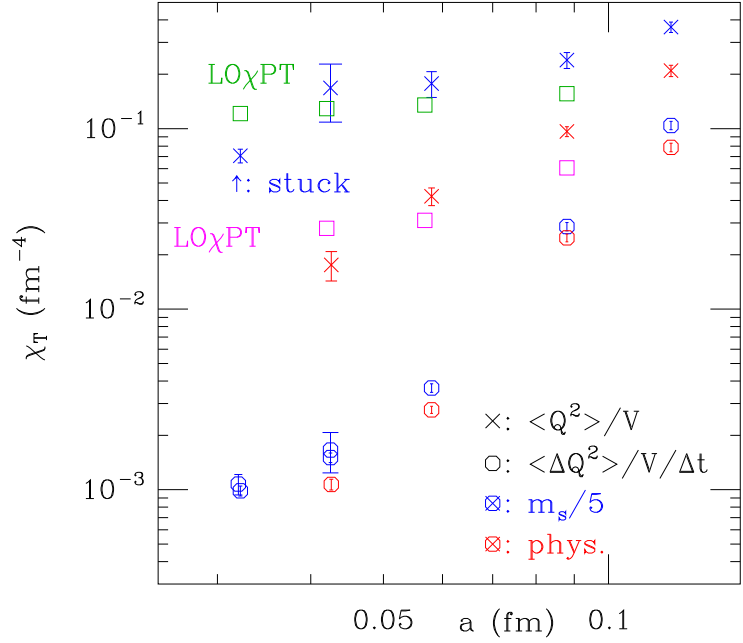}\end{center}
\vspace{-1.35in}
\caption{
\label{qsq_vs_a_fig}
Average topological susceptibility $\langle Q^2/V\rangle $ (crosses) and tunneling rate per unit volume
 $\langle \LP \Delta Q \RP^2/\LP V \Delta t \RP \rangle$ 
 (octagons) versus lattice spacing. Red and blue points are results for ensembles with $m_l=m_s/5$ and $m_l$ approximately physical, respectively.
Similarly, the magenta and green squares are the lowest order chiral perturbation theory
predictions for the susceptibility at $m_l=m_s/5$ and $m_l$ at its physical value, $\approx\! m_s/27$.
The leading-order \chpt\ results shown include staggered corrections and are taken from \eq{chiT-stag} below.}
\end{figure}

We define $\Delta Q$ as the change in $Q$ over molecular dynamics time $\Delta t$.
Figure~\ref{qsq_vs_a_fig} shows the tunneling rate per volume, $\langle \LP \Delta Q \RP^2/\LP V \Delta t \RP \rangle$ 
with octagons, where the blue symbols are for the $m_s/5$ ensembles and the red for the physical
$m_l$ ensembles.  The tunneling rate does not depend strongly on the quark
mass, but decreases as expected as the lattice spacing gets small. (In the cases where there are two blue
octagons, there were two sub-ensembles with a different molecular dynamics trajectory lengths.)
The crosses in Fig.~\ref{qsq_vs_a_fig} show the topological susceptibility, $\langle Q^2/V \rangle$.
Here we see the expected strong dependence on light quark mass.  The small error bar
on the $0.03$ fm point is unrealistic --- it simply reflects the fact that $Q$ is basically stuck
near this value in this simulation.

\section{Quantities at Fixed $Q$}
\label{sec:fixedQ}

In this section, we outline the relation between the behavior  of physical quantities at fixed
 topological charge and their dependence on the vacuum angle $\theta$.  The discussion relies heavily on
that  in \rcite{brower2003}.
For nonzero  $\theta$, the partition function is
\BNE \eqn{Ztheta}
Z(\theta) = \int {\cal D}A {\cal D}\bar\Psi {\cal D}\Psi \
\exp(- S[A,\bar\Psi,\Psi]) \exp(- i \theta Q[A]).
\ENE
The topological susceptibility $\chi_T$ is defined by \cite{witten1979,veneziano1979,leutwylersmilga1992}
\BNE
\chi_T \equiv -\frac{1}{V} \; \left(\frac{1}{Z}\frac{\partial^2
Z}{\partial\theta^2}\right)\Bigg\vert_{\theta=0}  = \ \frac{1}{V} \langle Q^2\rangle \ \ .\eqn{chi}
\ENE
We assume that both the time extent $T$ of our system and the 3-dimensional volume $V_3$ are large, so  the 4-dimensional volume $V=TV_3$ is also large. The partition function is then dominated by
the vacuum energy density  $\epsilon_0(\theta)$, 
\BEA\eqn{Z-eps}
Z(\theta) &\approx&  C\exp(-V\epsilon_0(\theta)),\\
\epsilon_0 &=& \frac{1}{2}\chi_T\theta^2 +\gamma\theta^4+\cdots,\eqn{eps0}
\EEAN  
where $C$ is a constant. The fact that $\chi_T$ is the coefficient of the quadratic term follows from \eq{chi}.
Parity symmetry (or, more precisely, {\it extended parity} --- see below) implies that only even powers of $\theta$ appear in \eq{eps0}.

Quantities evaluated at fixed $Q$ can be found by Fourier transforming 
\BEA \eqn{ZQ} Z_Q &=&\frac{1}{2 \pi} \int_{-\pi}^\pi d\theta \ \exp(i \theta Q) Z(\theta), \\
G_Q &=& \langle {\cal O}_1 {\cal O}_2 ... {\cal O}_n \rangle_Q =
\frac{1}{Z_Q} \frac{1}{2 \pi} \int_{- \pi}^\pi d\theta \
  \exp(i \theta Q) Z(\theta) G(\theta),  \eqn{GQ}
  \EEAN
with $G(\theta)= \langle {\cal O}_1 {\cal O}_2 ... {\cal O}_n \rangle_\theta$.
Since $V$ is large, we can do the $\theta$ integrals by the saddle point method.  Using \eqs{Z-eps}{eps0},
the saddle occurs at 
\BNE\eqn{thetas}
\theta_s =  i \frac{Q}{\chi_T V} +\cO\left(\frac{Q^3}{V^3}\right).
\ENE
This gives
\BNE G_Q = G(\theta_s) + \frac{1}{2\chi_TV } \PARTWO{G}{\theta}\big|_{\theta=\theta_s} + ... ,\eqn{GQ2}
\ENE
which in turn implies
 \cite{brower2003,aoki2007}
\BNE
\eqn{topo_dep}
B \big|_{Q,V} = B + \frac{1}{2\chi_T V} B''
   \LP 1-\frac{Q^2}{\chi_T V} \RP + {\cal O}\LP \frac{1}{\LP \chi_T V\RP^2} \RP, \ENE
where  $B$ is the mass $M$ or the decay constant $f$, and the primes here and below indicate derivatives with respect to $\theta$ evaluated at
$\theta=0$.  The terms $B$ and $-(B''/2)(Q/\chi_TV)^2$ on the right hand side of \eq{topo_dep}
come from expanding $G(\theta_s)$ in \eq{GQ2}, 
while the term $B''/(2\chi_TV)$ comes from the
term proportional to $\partial^2G/\partial\theta^2$.  By \eq{chi},
the correction to $B$ vanishes when averaged over $Q$.  \EQuations{GQ}{GQ2}{topo_dep} are
valid under
the assumption $Q \sim \sqrt{\chi_TV}$, \ie a ``typical'' value in a random-walk
of $Q$ around $Q=0$ with $\langle Q^2\rangle=\chi_T V$.

Here and below we use the fact that $B'=0$. This is true for any
parity-conserving quantity,  since, although parity symmetry is broken
at the QCD level by the $\theta$ term, 
what we can call {\it extended parity}\/ is preserved.  In extended parity,
we take $\theta\to-\theta$ along with a 
normal parity transformation.  This means that a matrix element that does not 
violate parity at $\theta=0$ must be even in $\theta$, and 
therefore its first derivative at $\theta=0$ vanishes.
For masses and decay constants, the \chpt\ 
results will confirm this.

\section{Chiral Perturbation Theory: Unitary Case} \label{sec:unitary}
The quantity $B''$ ($B= M$ or $f$) appearing in \eq{topo_dep} is physical and therefore has dependence on the
3-dimensional volume $V_3 = L^3$ that is exponentially suppressed, $\propto \exp(-M_\pi L)$.  This is in contrast with the
quantity at fixed $Q$, $B\vert_{Q,V}$, which has power-law dependence on $1/V$.    We may thus get a handle on topological
effects by calculating $B''$ in infinite-volume \chpt.  For now, we ignore possible discretization errors and consider
\chpt\ in the continuum only.  Corrections due to discretization effects for staggered quarks are calculated in
\secref{staggered}.  For two dynamical flavors, a first calculation of $M''$ in \chpt\ for full (unitary) QCD in infinite
volume and the continuum appears in \rcite{brower2003}.

In the presence of a vacuum angle $\theta$, the leading order (LO) Euclidean chiral 
Lagrangian is
\begin{equation}
\eqn{ChPT-L}
\cL_{\chi}= \frac{f^2}{8}\tr(\partial_\mu\Sigma\partial_\mu\Sigma^\dagger)
-\frac{B_0f^2}{4}\tr(\cM_A^*\Sigma + \cM_A\Sigma^\dagger),
\end{equation}
where the normalization is such that $f\approx 130$ MeV, and
where $\cM_A\equiv e^{i\theta/N} \cM$, with $N$ the number of flavors and $\cM$ the mass matrix in the absence of
$\theta$.   Complex conjugation is denoted by $*$.  We always take $\cM$ to be diagonal in this paper. 
The change from $\cM$ to $\cM_A$ is effected by an anomalous (flavor-singlet)
chiral transformation, which simultaneously removes the $i\theta Q$ from the
Euclidean
gauge action.    See Appendix~\ref{m=0} for more discussion of the phases in the mass matrix.

In order to set out notation and make various points that will be useful later, we examine
both the two-flavor and the three-flavor unitary cases in detail.  We emphasize that none of the results
in this section are new to the literature; while  \rcite{brower2003} does not discuss decay constants
or the $N=3$ case, those results can easily be obtained as limits of the general partially quenched results
given in \rcite{aokifukaya2009}.

\subsection{Two flavors}
\label{sec:nf2}

We consider $N=2$ case  with nondegenerate 
quark masses $m_u\not= m_d$.  With $\theta\not=0$, $\Sigma$ can have 
a nontrivial vacuum expectation value, $\langle\Sigma\rangle$.  
When the quark masses are nondegenerate, an  argument for arbitrary $N$ by Gasser and
Leutwyler \cite{GandL} shows that
  $\langle\Sigma\rangle$ must be diagonal.\footnote{If there are
  degeneracies, then $\langle\Sigma\rangle$ can be put in diagonal form by making a vector (flavor) rotation
  that leaves the chiral Lagrangian unchanged.}
    The intuitive reason for this is that  only the diagonal elements
  of $\Sigma$ enter into the potential energy, which has an overall negative sign in \eq{ChPT-L}. Since $\Sigma$ is unitary, off-diagonal
  elements would reduce the absolute size of the diagonal elements and result in a higher potential energy.  

We therefore let
\begin{equation}
\langle \Sigma\rangle = 
\begin{pmatrix} e^{i\alpha} & 0 \\ 0 & e^{-i\alpha}
\end{pmatrix},
\end{equation}
where the diagonal elements are constrained by $\det{\Sigma}=1$.
The potential energy term we need to minimize is then
\begin{equation}
\eqn{energy-nf2}
V= -\frac{B_0f^2}{2}\Big(m_u \cos(\alpha-\theta/2)+m_d\cos(\alpha+\theta/2)\Big). 
\end{equation}
Differentiating with respect to $\alpha$ gives the condition
\begin{equation}
\eqn{condition-nf2}
m_u \sin(\alpha-\theta/2)+m_d\sin(\alpha+\theta/2) = 0,
\end{equation}
with the solution
\begin{equation}
\eqn{solution-nf2}
\tan(\alpha)= r \tan(\theta/2), \qquad r \equiv \frac{m_u-m_d}{m_u+m_d}.
\end{equation}

We can now expand the potential energy to quadratic order to find the 
pion mass. The meson field $\Phi$, which
characterizes fluctuations around the vacuum expectation value, is defined most conveniently by
\begin{eqnarray}
\Sigma &=& s\; e^{2i\Phi/f}\; s, \qquad\quad s \equiv \sqrt{\langle \Sigma\rangle} 
\eqn{Sigma} \\
\Sigma^\dagger &=& s^\dagger\; e^{-2i\Phi/f}\; s^\dagger. 
\eqn{Sigmadagger} 
\end{eqnarray}
The field $\Phi$ may be written as usual in terms of individual meson fields as
\begin{equation}
\Phi=
\begin{pmatrix} \pi^0/\sqrt{2} & \pi^+ \\ \pi^- & -\pi^0/\sqrt{2} 
\end{pmatrix}.
\eqn{Phi2}
\end{equation}

The definition of $\Phi$ in \eq{Sigma} is convenient for two reasons.  First of all, it transforms normally
under extended parity 
(usual parity plus $\theta\to-\theta$):
\begin{eqnarray}
\eqn{Phi-parity}
\Phi &\to& - \Phi ,\\  
\eqn{s-parity}
s &\to& s^\dagger  ,\\  
\eqn{Sigma-parity}
\Sigma &\to& \Sigma^\dagger.
\end{eqnarray}
This means that the fields $\pi^{\pm},\pi^0$ have the usual interpretation as pion fields.  (Note that
\eq{s-parity} follows from the fact that $\alpha\to-\alpha$ when
$\theta\to-\theta$.) The chiral Lagrangian, \eq{ChPT-L}, is easily seen to be invariant
under extended parity, as expected. Secondly, with the definition \eq{Sigma}, the kinetic energy term 
in \eq{ChPT-L} takes the same form in terms of $\Phi$ as is does in the standard case when
$\langle\Sigma\rangle=I$. This means that there is no wave function renormalization at leading order, which
simplifies calculations.  An alternative definition,
\begin{equation}
\Sigma = e^{i\Phi/f}\;\langle\Sigma\rangle\; e^{i\Phi/f}\;\eqn{alt-Sigma},
\end{equation}
also leads to \eqs{Phi-parity}{Sigma-parity} under extended parity, but generates a nontrivial leading-order 
wave function renormalization factor.  Of course, physical results must be the same with any 
appropriate field definition.  It is straightforward to check that the  meson mass and decay constant results
below, found in the first instance using \eq{Sigma}, can also be obtained with \eq{alt-Sigma}.

Expanding to quadratic order in $\Phi$,
 we find, for the charged pion mass,
\begin{equation}
\eqn{mpi-nf2}
M_\pi^2(\theta) = M_\pi^2(0)\cos\left(\frac{\theta}{2}\right)\sqrt{1+r^2\tan^2
\left(\frac{\theta}{2}\right)},
\end{equation}
where $M_\pi^2(0) = B_0(m_u+m_d)$. \Equation{mpi-nf2}
agrees with \rcite{brower2003}, Eq.\ (4.15).

For the decay constant we need the axial current $\cA_\mu^{ij}$ in \chpt\ 
that corresponds to the QCD current $\bar q^j \gamma_\mu\gamma_5 q^i$,
where $i,j$ are flavor indices.  With the definitions of \rcite{GandL},
\begin{equation}
\cA_\mu^{ij} = i\frac{f^2}{4}\left(\partial_\mu\Sigma\Sigma^\dagger +
\Sigma^\dagger \partial_\mu\Sigma\right)^{ij}.
\eqn{axial-current}
\end{equation}
Note that the axial current comes from the kinetic energy term in the chiral
Lagrangian, and its form in terms of $\Sigma$ is unaffected by a nonzero $\theta$.  
For $f_\pi$, we need $\cA_\mu^{12}$.
Plugging in \eq{Sigma}, gives, to leading order, 
\begin{equation}
\cA_\mu^{12} = -f\cos(\alpha)\partial_\mu \pi^+.
\eqn{axial-current-LO}
\end{equation}
With \eq{solution-nf2},
this implies
\begin{equation}
f_\pi(\theta)  = f\cos(\alpha) = f_\pi(0) \cos(\alpha)=
\frac{f_\pi(0)}{\sqrt{1+r^2\tan^2(\theta/2)}}. 
\eqn{fpi-nf2}
\end{equation}
Note that $f_\pi$ is independent of $\theta$ in the degenerate case, $r=0$.
To apply \eq{topo_dep}, we need the second derivative of $M_\pi$ or 
$f_\pi$ at $\theta=0$.  From \eqs{mpi-nf2}{fpi-nf2}, we obtain
\begin{eqnarray}
\eqn{mpi2pp}
M_\pi'' &=& M_\pi(0) \frac{r^2-1}{8} = - M_\pi(0)\; \frac{m_u m_d}{2(m_u+m_d)^2}, \\  
f_\pi'' &=& -f_\pi(0) \frac{r^2}{4} = -f_\pi(0)\; \frac{(m_u-m_d)^2}{4(m_u+m_d)^2}.
\eqn{fpipp}
\end{eqnarray}
As expected, the first derivatives, $M_\pi'$ 
and $f_\pi'$, vanish.  Note that $f_\pi''$ does
not vanish when one mass ($m_u$, say) goes to zero.  This seems to contradict the 
expectation that the theory is $\theta$-independent when one quark mass vanishes.  However, the
$\theta$-independence does not apply to quantities, such as  $f_\pi$, that depend on  external currents.
  We explain this in more detail in Appendix \ref{m=0}.

\subsection{Three flavors}
\label{sec:nf3}
For $N=3$, we will work in the limit $m_u=m_d\equiv m$, but $m\not=m_s$ in 
general. Since isospin is 
preserved, we can assume
\begin{equation}
\eqn{Sigma-nf3}
\langle \Sigma\rangle = 
\begin{pmatrix} e^{i\alpha} & 0 & 0 \\ 0 & e^{i\alpha} & 0 \\ 0 & 0 & e^{-2i\alpha}
\end{pmatrix}.
\end{equation}
The potential energy term is then
\begin{equation}
\eqn{energy-nf3}
V= -\frac{B_0f^2}{2}\Big(2 m \cos(\alpha-\theta/3)+m_s\cos(2\alpha+\theta/3)\Big). 
\end{equation}
Differentiating with respect to $\alpha$ gives the condition
\begin{equation}
\eqn{condition-nf3}
2 m \sin(\alpha-\theta/3)+2 m_s\sin(2\alpha+\theta/3) = 0.
\end{equation}

Although \eq{condition-nf3} does not have a simple analytic solution, we
really only need derivatives of quantities at $\theta=0$, which
can all be calculated by implicit differentiation.  Note first
that the solution for $\langle \Sigma \rangle$ is invariant under
$\theta\to-\theta$, $\alpha\to-\alpha$, so the solution 
$\alpha(\theta)$ has only odd powers of $\theta$.  In particular, the second
derivative of $\alpha(\theta)$ at $\theta=0$ vanishes: $\alpha''=0$.
We then write a physical quantity $W(\theta)$ as $W(\alpha(\theta),\theta)$, where the second argument is the explicit $\theta$-dependence, and the 
first is the dependence through $\alpha$.  Using $\alpha''=0$, we have
\begin{equation}
W''\equiv\frac{d^2 W}{d\theta^2}\Big\vert_{\theta=0} =
\frac{\partial^2 W}{\partial \alpha^2}(\alpha')^2 +
2\frac{\partial^2 W}{\partial \alpha\partial\theta}\;\alpha' +
\frac{\partial^2 W}{\partial \theta^2}\ ,
\eqn{Wpp}
\end{equation}
where all derivatives on the right-hand side are to be evaluated at $\theta=0$
(which implies $\alpha=0$).  Thus, all we need from the solution 
to \eq{condition-nf3} is $\alpha'$, which is easily calculated to be
\begin{equation}
\eqn{alphap-nf3}
\alpha'= \frac{m-m_s}{3(m+2m_s)}.
\end{equation}

We now merely need to find the masses and decay constants as functions
of $\alpha$ and $\theta$.  Expanding the potential energy term in \eq{ChPT-L}
to quadratic order in the now $3\times3$ meson matrix $\Phi$, we find
\begin{eqnarray}
M_\pi^2 &=& B_0\Big(2m\cos(\alpha-\theta/3)\Big),\nonumber \\
M_K^2 &=& B_0\Big(m\cos(\alpha-\theta/3)+m_s\cos(2\alpha+\theta/3)\Big).
\eqn{masses-nf3}
\end{eqnarray}
Similarly, using \eq{Sigma} and 
expanding \eq{axial-current} to linear order in $\Phi$ gives 
\begin{equation}
f_\pi = f, \qquad f_K = f\cos(3\alpha/2).
\eqn{decayconsts-nf3}
\end{equation}
Using \eq{Wpp} now gives
\begin{eqnarray}
M_\pi'' &=& -M_\pi(0)\;\frac{m^2_s}{2(m+2m_s)^2}, \eqn{mpi-nf3}\\
M_K'' &=&  -M_K(0)\;\frac{m m_s}{2(m+2m_s)^2}, \eqn{mK-nf3}\\
f_\pi'' &=& 0,
\eqn{fpi-nf3}\\
f_K'' &=& -f_K(0)\; \frac{(m_s-m)^2}{4(m+2m_s)^2}.
\eqn{fK-nf3}
\end{eqnarray}
Like $f''_\pi$ in the  two-flavor case, $f_K''$ does
not vanish when one of the masses goes to zero;
 see Appendix \ref{m=0} for an explanation.

\section{Partially Quenched \chpt}
\label{sec:PQ}

Since most of our lattice data is partially quenched, we need to extend the
calculation 
of $M''$ and $f''$ to partially quenched \chpt\ (PQ\chpt).
This was done by Aoki and Fukaya~\cite{aokifukaya2009} 
using the replica method to remove the determinant of the valence quarks.
However, the required calculation is nonperturbative, at least on its face, since  the vacuum state changes 
in the presence of $\theta$.  The replica method is only justified perturbatively, so a nonperturbatively safe
method is preferable. The
Lagrangian approach of \rcite{bernardgolterman1993}, which introduces ghost (bosonic) quarks to 
cancel the valence quark determinant, is also only valid perturbatively, since it ignores the 
requirement that the
bosonic path integral be convergent.  

The nonperturbatively correct version of \pqchpt\ has been worked out by
Golterman, Sharpe, and Singleton \cite{gss} and Sharpe and Shoresh \cite{sharpeshoresh2001}. 
The nonperturbative problems of the Lagrangian approach are fixed by taking into account the convergence requirement.

In terms of $\Sigma$, the chiral Lagrangian is
\begin{equation}
\eqn{PQChPT-L}
\cL_{\chi,PQ}= \frac{f^2}{8}\str(\partial_\mu\Sigma\partial_\mu\Sigma^{-1})
-\frac{B_0f^2}{4}\str(\cM\Sigma + \cM\Sigma^{-1})\ ,
\end{equation} 
with \str\ the supertrace.  The main difference from the standard perturbative version of
\pqchpt\ \cite{bernardgolterman1993} is that $\Sigma$ is not unitary, which is why 
$\Sigma^{-1}$ appears instead of $\Sigma^\dagger$ in \eq{PQChPT-L}.  

For definiteness,
we work with three sea quarks ($N=3$) and two valence quarks ($N_v=2$), and
take the isospin limit in the sea:  $m_u=m_d\equiv m$, 
but $m\not=m_s$ in general. The  valence quarks are $x$ and $y$ with
masses $m_x$ and $m_y$, respectively.  Corresponding ghost quarks $\tilde x$ and $\tilde y$ with
masses $m_x$ and $m_y$
are included to cancel the valence-quark determinant. The chiral field $\Sigma$ and the quark mass
matrix $\cM$ are then
 $7\times7$ matrices.  The mass matrix is given by
\begin{equation}\eqn{PQ-mass-matrix}
\cM={\rm diag}(m,m,m_s,m_x,m_y,m_x,m_y),
\end{equation}
where the quarks are ordered: sea, valence, ghost.

In expanding $\Sigma$ in terms of pseudoscalar meson fields, 
it is useful 
to separate out a special diagonal meson field, $\epsilon$, which is a linear combination of flavor-neutral quark-antiquark and ghost-antighost mesons.  We write
\begin{eqnarray}
\Sigma &=& \exp(2i\Phi/f)\eqn{PQSigma},\\  
\Phi &=& \Phi' + i\epsilon T_6 \eqn{Phi},\\ 
\Phi' &=& 
\begin{pmatrix} \phi & \eta \\ \eta^\dagger & i\tilde\phi
\end{pmatrix} , \eqn{Phip}
\end{eqnarray}
where $\phi$ is the quark-antiquark block (both sea and valence), $\tilde \phi$ is the ghost-antighost
block, and $\eta$ and $\eta^\dagger$ are the quark-antighost and ghost-antiquark blocks, respectively. The
diagonal generator%
\footnote{We follow the notation of \rcite{sharpeshoresh2001} in naming the diagonal generators.}
$T_6$ is non-anomalous (straceless), and is given by
\begin{equation}\eqn{T6}
T_6=  \sqrt{\frac{2}{15}}\,{\rm diag}(1,1,1,1,1,5/2,5/2).
\end{equation}
 Both $\phi$ and $\tilde \phi$ are Hermitian and traceless.  The factor of $i$ in the 
ghost-antighost ($\tilde\phi$) block in \eq{Phip} comes ultimately from the careful consideration of the true symmetries of the theory with ghosts. These
are more complicated than those assumed in \rcite{bernardgolterman1993} because of the necessity of keeping
the ghost (bosonic) path integrals convergent in a nonperturbative treatment. At the chiral level, 
the integrals over the independent real fields in $\tilde \phi$ run from $-\infty$ to $\infty$, and the 
factor of $i$ multiplying $\tilde\phi$ in \eq{Phip}, in combination with the supertrace, guarantees that the action for these fields is positive
definite, so the kinetic and mass terms have the proper sign for convergence of the path integral. 
There is no problem with the convergence of the $\phi$ and $\eta,\eta^\dagger$ integrals because the former are over a compact region (they are angles), and the latter are Grassmann variables.

In \eq{Phi}, we have followed the prescription of Sharpe and Shoresh
\cite{sharpeshoresh2001} and have included a factor of $i$ with the field $\epsilon$  multiplying $T_6$. 
Because $\str(T_6^2)< 0$,
the $i$ is  necessary in order for the kinetic energy of $\epsilon$  to be positive.  
In other words, $\epsilon$ is ``ghost-like,'' rather than ``quark-like.''  Like $\tilde\phi$, $\epsilon$ should be 
integrated along the entire real axis.  
  Note that the only other linearly independent diagonal 
generator to span the quark-antiquark and ghost-antighost blocks is 
\begin{equation}\eqn{T7}
T_7=  \frac{1}{\sqrt{3}} I= \frac{1}{\sqrt{3}}\,{\rm diag}(1,1,1,1,1,1,1) ,
\end{equation}
with $I$ the identity matrix.
Since $T_7$ is anomalous ($\str(T_7)=\sqrt{3}$), the corresponding meson 
(called $\Phi_0$, or less precisely, $\eta'$) is heavy and is 
integrated out of the chiral theory.

When including the $\theta$ angle, the most natural approach would be to remove 
 the $\theta F\tilde F$ term
by making an anomalous rotation using $T_7$ as generator.  At the chiral level, this would put a factor of
$\exp(-i\theta/3)$ in front of the $\str(\cM\Sigma)$ term in \eq{PQChPT-L}, and a factor of
$\exp(i\theta/3)$ in front of the $\str(\cM\Sigma^{-1})$ term.  In other words, all quarks (sea and valence)
and all ghosts would get the same $\theta$ phase.   However, it is convenient to make  an additional non-anomalous rotation with the generator
\begin{equation}\eqn{t}
t={\rm diag}(0,0,0,1,1,1,1),
\end{equation}
 which is a linear combination of $T_6$ and a generator in the quark-antiquark block, namely
${\rm diag}(1,1,1,-3/2,-3/2,0,0)$.
This allows us to remove the $\theta$-dependent phase from all mass terms of valence and ghost quarks,
which makes the algebra somewhat simpler and has further advantages for the heavy-light case discussed
 in \secref{heavy-light}.  The partially quenched chiral Lagrangian in the presence of $\theta$ is then
\begin{equation}
\eqn{PQChPT-Ltheta}
\cL_{\chi,PQ,\theta} = \frac{f^2}{8}\str(\partial_\mu\Sigma\partial_\mu\Sigma^{-1})
-\frac{B_0f^2}{4}\str(\cM_B^*\Sigma + \cM_B\Sigma^{-1})\ ,
\end{equation} 
where \begin{equation}
\eqn{MB}
\cM_B \equiv  {\rm diag}(e^{i\theta/3}m, e^{i\theta/3}m, e^{i\theta/3}m_s,m_x,m_y,m_x,m_y).
\end{equation} 
In Appendix \ref{m=0}, several choices for the mass matrix in the presence of $\theta$ are discussed;
we include the subscript $B$ on $\cM_B$ for consistency with notation introduced there.

In the unitary theory, the absolute minimum of the potential energy term determines  the vacuum state
$\langle \Sigma \rangle$.  Here, the potential energy $V$ is complex. \Rcite{gss} argues that  we should 
therefore find a saddle point  of $|\exp(-V)|= \exp(-\Re(V))$ (deforming the $\tilde\phi$ 
and $\epsilon$ contours 
as needed), not a minimum. 
One issue that arises in the saddle point analysis is how to choose the proper saddle point when the complex
saddle-point equation, $V'=0$ has multiple solutions, as it does here.
 Fortunately, \rcite{deBruijn} (referred to
by \rcite{gss}) gives a prescription for finding the unique useful saddle point for an analytic function like $V$:
Find (1), a point that is a solution of $V'=0$, and (2), a deformation of the contour that goes through 
the point in the direction of steepest descent of  $|\exp(-V)|$ (steepest ascent of $\Re(V)$) and satisfies
the requirement that the  point has the highest value of
$|\exp(-V)|$ of any place on the deformed contour.  There is at most one saddle point that satisfies these 
conditions, so once we find one such point, we are guaranteed to have found the unique solution, which
determines $\langle\Sigma\rangle$.

The problem of solving $V'=0$ is simplified by noting that, as in the unitary theory, 
$\langle \Sigma \rangle$ is diagonal.  This can be proved by following the unitary-theory argument in \rcite{GandL}.
For the case when all masses are nondegenerate, the argument goes through with only trivial modifications.
Degeneracies among sea quarks, or between valence and sea quarks, also present no problem because
$\Sigma$ is a unitary matrix with $c$-number entries in these blocks, and can be diagonalized exactly as in
the unitary theory.  
However, degeneracy between valence and ghost quarks 
must be considered because such degeneracies are built into the partially
quenched theory. These degeneracies are different than those among quarks since the graded structure of
the group is crucial.  It is plausible  that the block of $\langle \Sigma \rangle$ corresponding
to the degenerate mass pair can be diagonalized by a vector similarity transformation, $\Sigma\to U\Sigma
U^{-1}$, where $U$ is an element of the graded symmetry group $SL(N+N_v|N_v)$ (\ie $SL(5|2)$ here). 
Such a transformation leaves the Lagrangian, including the mass (potential energy) term, unchanged.  We have
checked the diagonalization explicitly for the 
crucial 2-fold degeneracy of quark and ghost, and believe it must also be true if
there is a higher degeneracy (\eg $m_x=m_y$ also), but have not proved it.

Following
\eqsthru{PQSigma}{Phip}, we therefore
parameterize $\langle\Sigma\rangle$ as
\begin{eqnarray}\eqn{vacuum-general}
\langle\Sigma\rangle &=&\exp\Big(i\,{\rm diag}(\alpha+\delta+i\epsilon,\;\alpha+\delta+i\epsilon,\;
-2\alpha+\delta+i\epsilon,\; \beta-3\delta/2+i\epsilon,\;\nonumber\\
&&\hspace{20mm}-\beta-3\delta/2+i\epsilon,\;
 5i\epsilon/2-i\gamma,\;5i\epsilon/2+i\gamma)\Big),
\end{eqnarray}
where we have ensured that the exponent is straceless, and have used isospin symmetry 
($m_u=m_d\equiv m$) to require that the first two entries along the diagonal be equal.  We have chosen
simpler normalization for the angles in \eq{vacuum-general} than we would need to use for
the corresponding meson fields.  

Since we will deform the contour for the ghost-antighost fields, the 
variables $\gamma$ and $\epsilon$ may be complex.
For convenience we define $\epsilon \equiv i\hat\epsilon$, $\gamma=i\hat\gamma$, where it will turn out
that $\hat\gamma$ and $\hat\epsilon$ are in fact real at the saddle point. With this definition, the potential
energy is
\begin{eqnarray}\eqn{V-general}
V &=&-\frac{B_0f^2}{2}\bigg\{2m\cos(\theta/3-\alpha-\delta+\hat\epsilon)
+m_s\cos(\theta/3+2\alpha-\delta+\hat\epsilon)+ \nonumber\\
&&\hspace{20mm}+m_x\cos(3\delta/2- \beta+\hat\epsilon)+ m_y \cos(3\delta/2+\beta+\hat\epsilon)+\nonumber\\ &&\hspace{20mm}-m_x\cos(5\hat\epsilon/2-\hat\gamma)-m_y\cos(5\hat\epsilon/2+\hat\gamma)\bigg\}.
\end{eqnarray} 
From the requirement that $V$ is stationary at the saddle point with respect to $\alpha$, $\beta$, $\hat\gamma$,
$\delta$ and $\hat\epsilon$, respectively, we  obtain the equations:
\begin{eqnarray}\eqn{stationary-point}
&&m\sin(\theta/3-\alpha-\delta+\hat\epsilon) -m_s\sin(\theta/3+2\alpha-\delta+\hat\epsilon)  =  0, \eqn{alpha-eq}\\
&&m_x\sin(3\delta/2-\beta+\hat\epsilon) -m_y\sin(3\delta/2+\beta+\hat\epsilon)  =  0, \eqn{beta-eq} \\
&&m_x\sin(5\hat\epsilon/2-\hat\gamma) -m_y\sin(\hat\gamma+5\hat\epsilon/2)  =  0,\eqn{gamma-eq}\\
&&2m\sin(\theta/3-\alpha-\delta+\hat\epsilon) +m_s\sin(\theta/3+2\alpha-\delta+\hat\epsilon) + \hspace{10mm}  \nonumber\\
&&\hspace{5mm}-\frac{3}{2}m_x\sin(3\delta/2-\beta+\hat\epsilon) -\frac{3}{2}m_y\sin(3\delta/2+\beta+\hat\epsilon)  =  0 ,\eqn{delta-eq}\\
&&2m\sin(\theta/3-\alpha-\delta+\hat\epsilon) +m_s\sin(\theta/3+2\alpha-\delta+\hat\epsilon)  +  \nonumber\\
&&\hspace{5mm}+m_x\sin(3\delta/2-\beta+\hat\epsilon) +m_y\sin(3\delta/2+\beta+\hat\epsilon)+\nonumber\\
&&\hspace{10mm}-\frac{5}{2}m_x\sin(5\hat\epsilon/2-\hat\gamma) -\frac{5}{2}m_y\sin(\hat\gamma+5\hat\epsilon/2)  =  0 \eqn{epsilon-eq}.
\end{eqnarray}

In the case $\theta=0$, we have the standard perturbative solution:  $\alpha=\beta=\hat\gamma=\delta=\hat
\epsilon=0$, so $\langle\Sigma\rangle=I$.  If the valence masses $m_x$ and $m_y$
are not too small,  it is easy to check that this saddle point is the
correct one to use because $\Re(V)$ increases
monotonically away from the saddle on the original
contours, on which $\epsilon$ and $\gamma$ are real.   
However, Sharpe and Shoresh  \cite{sharpeshoresh2001} found a lower bound on the valence
masses, below which  the 
real part of the squared-mass matrix of the ghost-like neutral fields is not positive definite.  When the valence masses violate the bound,  $\Re(V)$  decreases 
away from the saddle in a
real direction on one of the contours.  At first glance, this suggests that the perturbative
vacuum is not the correct one in this case.  We discuss the issue in detail
in Appendix \ref{ghost-mass}, and show that the monotonic increase away from the perturbative
saddle point  is restored after one or more of the neutral quark-antiquark integrals are performed.
This means that we may freely violate the Sharpe-Shoresh bound, and the perturbative
vacuum is the correct one for any nonzero values of the valence-quark masses.

For $\theta$ not too far from 0, we  expect that
the proper saddle point  is  then the nearby one, where the
magnitudes of the arguments of all the sine functions are less than $\pi/2$. If this were not true, it
would invalidate the analysis that led to \eq{topo_dep}, since we assumed a smooth dependence on $\theta$.
Nevertheless, to be sure our analysis is correct nonperturbatively, we will check this assumption
below.  

With the arguments bounded by assumption,  two sine functions are equal if and only if their arguments are equal.
Subtracting \eq{delta-eq} from \eq{epsilon-eq} to eliminate terms with $m$ and $m_s$ and then using \eqs{beta-eq}{gamma-eq} to eliminate terms with $m_x$ or $m_y$ then implies $\hat\epsilon=\delta$ and
$\hat\gamma=\beta$.  The saddle point value $\langle\Sigma\rangle$ simplifies to
\begin{eqnarray}\eqn{vacuum}
\langle\Sigma\rangle &=&\exp\Big(i\,{\rm diag}(\alpha,\;\alpha,\;
-2\alpha,\; \beta-5\delta/2,\;-\beta-5\delta/2,\;\nonumber\\
&&\hspace{20mm}
 \beta-5\delta/2,\;-\beta-5\delta/2)\Big).
\end{eqnarray}
This has two required features of a partially quenched theory:  (1) the sea-quark sector is unaffected by the 
presence of valence quarks and ghosts, and (2) the vacuum expectation values of $\bar q q$ and $\bar g g$ 
are equal for corresponding quark ($q$) and ghost ($g$).

Plugging the results for $\hat \epsilon$ and $\hat \gamma$ into the saddle point equations, 
\eqsthru{alpha-eq}{epsilon-eq}, then determines the remaining variables, $\alpha$, $\beta$ and $\delta$.  It is 
 not  necessary to obtain a closed-form 
 solution.  As in
the unitary 3-flavor case, we only  need $\alpha'$, $\beta'$ and $\delta'$, the derivatives 
of these angles 
with respect to $\theta$ at $\theta=0$.  By differentiating the saddle point equations and solving, we find
\begin{eqnarray}
\eqn{alphap-again}
\alpha' &=& \frac{m-m_s}{3(m+2m_s)} \ ,\\
\eqn{betap}
\beta' &=& \frac{m_x-m_y}{2m_xm_y}\cdot\frac{m m_s}{m+2m_s}=\hat\gamma'\ , \\
\eqn{deltap}
\delta' &=& \frac{m_x+m_y}{5m_xm_y}\cdot\frac{m m_s}{m+2m_s}=\hat\epsilon'\ .
\end{eqnarray}
As expected, the angle governing the sea-quark vacuum expectation value, $\alpha$, 
obeys the same equation as in the unitary QCD case, \eq{alphap-nf3}.  

Before proceeding, we should check that the saddle we have found is the proper one to use.   At the saddle
point, $\hat\epsilon$ and $\hat\gamma$ (the imaginary parts of $\epsilon$ and $\gamma$) are equal to $\delta$ and $\beta$ respectively, which are comparable to $\theta$ and hence small angles. 
If the Sharpe-Shoresh bound is satisfied, the steepest descent directions, in which $\Re(V)$ increases most rapidly away from the saddle point, are the real directions
for $\epsilon$ and $\gamma$.  As we continue the contours for
$\epsilon$ and $\gamma$ in these directions,
$\Re(V)$ increases exponentially,  dominated by one or both of the  ghost terms that
grow like $\cosh(\Re(5\epsilon/2\pm\gamma))$.  Far from the saddle point, it is then straightforward to see that
we can bend the contours back to the real axis
while keeping $\Re(V)$ large, \ie much larger than at the saddle point.  We  have therefore found a proper saddle 
point and contours.  Appendix \ref{ghost-mass} argues that the saddle and contours are still the correct 
ones when the
Sharpe-Shoresh bound is violated.

Using  \eqsthru{alphap-again}{deltap} to calculate the valence-meson mass and decay constant as
in \secref{nf3}, we find:
\begin{eqnarray}
M_{xy}'' &=& -M_{xy}(0)\;\frac{m^2 m^2_s}{(m+2m_s)^2}\;\frac{1}{2m_x m_y},
\eqn{mxy-nf3}\\
f_{xy}'' &=&   -f_{xy}(0)\; \frac{m^2m_s^2}{(m+2m_s)^2}\;\frac{(m_x-m_y)^2}{4m^2_x m^2_y} .
\eqn{fxy-nf3}
\end{eqnarray}
The results for the unitary pion and kaon, \eqsfour{mpi-nf3}{mK-nf3}{fpi-nf3}{fK-nf3},
can be obtained from \eqs{mxy-nf3}{fxy-nf3} in the appropriate limits:  $m_x=m_y=m$ for
the pion and $m_x=m$, $m_y=m_s$ for the kaon.

\Equations{mxy-nf3}{fxy-nf3} have singular limits when $m_x$ or $m_y$ or both go to zero at fixed
sea quark masses.  (For $f_{xy}$ one has to keep $m_x\not=m_y$ if both go to zero to get a 
nonzero result.) Such mass singularities are typical for partially quenched theory, but this is
the only case we know of where they appear at tree level.
On the other hand, the results  vanish when either sea-quark mass goes to zero, as explained in Appendix \ref{m=0}.

The results in \eqs{mxy-nf3}{fxy-nf3} agree with those computed by Aoki and Fukaya \cite{aokifukaya2009}, 
who used the
replica method for the partially quenched theory.  Because the replica method has not 
been nonperturbatively justified, the methods of 
\rcite{gss,sharpeshoresh2001} seem preferable to us here, since the ground
state of the theory is changing.  The agreement of the two methods suggests, though, that this particular
problem is essentially perturbative.  This makes sense because we in the end we only need derivatives of
quantities at $\theta=0$ --- so the dependence on the $\theta$ is required only in an
infinitesimal neighborhood of the perturbative  vacuum.   It has been a surprise to us
that the more vexing nonperturbative  issue in our analysis arises from the Sharpe-Shoresh bound, 
which  already affects the $\theta=0$ case.

\section{Heavy-Light Mesons}
\label{sec:heavy-light}
We now add a heavy quark $Q$ to the theory.  It is useful to consider the heavy quark in a partially-quenched
context:  let its valence mass be $m_Q$ and its sea mass be $ m_{Q,{\rm sea}}$.  In the presence of 
a nonzero $\theta$, we  put the $\theta$-dependence into the sea-quark 
(but not the valence-quark) mass matrix, as  in $\cM_B$, \eq{MB}. As both $m_Q$ and 
$ m_{Q,{\rm sea}}$ get large, the heavy sea quark decouples, and we are left with a theory of
light sea-quarks only.  The valence heavy quark of course does not decouple, since it can appear in
external states, but it carries no $\theta$-dependent phases. Note that it does not make any physical difference 
how the $\theta$-dependence is put into initial sea-quark mass matrix, \ie whether or not the heavy sea-quark
mass carries 
$\theta$-dependence.   The end result after  
$m_{Q,{\rm sea}}\to\infty$ is always the same as if we had started with a theory of only light sea quarks.
However the decoupling is indeed more subtle when the heavy 
sea-quark carries  
$\theta$-dependence ---
we cannot simply delete the heavy sea-quark terms from the Lagrangian. See  Appendix \ref{decoupling}
for a discussion of how decoupling works in that case.

The leading-order heavy-meson chiral Lagrangian
is then exactly the standard one \cite{MAN_WISE}:
\begin{equation}\label{eq:LHL}
  \cL_{\chi,{\rm HL}}  =  -i\, \sTr(\overline{H} H v\negcdot \leftvec D )
  + g_\pi\, \sTr(\overline{H}H\gamma^{\mu}\gamma_5
  \mathbb{A}_{\mu}) \ ,
\end{equation}
where
$H$ is the heavy-light meson field, composed of a pseudoscalar meson $P$ and a vector meson $P^*$:
\begin{equation} \label{eq:H_definition}
  H_{a} = \frac{1 + \vslash}{2}\left[ \gamma^\mu P^{*}_{\mu a}
    + i \gamma_5 P_{a}\right]\ ,
\end{equation}
with $v$ the meson's velocity, and
$a$ the flavor index of the light quark.
In \eq{LHL}, $\leftvec D$ is the covariant derivative (acting to the left), $\sTr$ is a trace over Dirac indices and a supertrace over flavor indices,%
\footnote{The supertrace is used because the theory is partially quenched. In most cases, however, the difference between trace and supertrace is irrelevant for the heavy-light part of the Lagrangian, since
closed heavy-light meson loops are forbidden anyway.} and $\mathbb{A_\mu}$ is the 
light-quark axial current,
\begin{eqnarray}
   \mathbb{A}_{\mu} & = & \frac{i}{2} \left[ \sigma^{\dagger} \partial_\mu
   \sigma - \sigma \partial_\mu \sigma^{\dagger}   \right] \ , \\
   \sigma &\equiv&  \sqrt{\Sigma}\ .
\end{eqnarray}

The leading-order left-handed current that destroys a heavy-light meson with light flavor $b$ is \cite{MAN_WISE}
\begin{equation}\label{eq:left-current}
  j_L^{\mu,b} = \frac{\kappa}{2}\;
  \trD\bigl(\gamma^\mu\left(1-\gamma_5\right)H\bigr) \sigma^\dagger\lambda^{(b)} ,  
  \end{equation}
where $\kappa$ is a low-energy constant, $\trD$ is a trace over Dirac indices only, and $\lambda^{(b)}$ is a constant column vector 
that fixes the flavor of the light quark:
$(\lambda^{(b)})_a = \delta_{a b}$.  For the decay constant, we need the heavy-light axial current 
\begin{equation}\label{eq:HQaxial-current}
  j_5^{\mu,b} =   \frac{1}{2}\!\left(j_R^{\mu,b}- j_L^{\mu,b}\right) = \frac{\kappa}{4}\left[
  \trD\bigl(\gamma^\mu\left(1+\gamma_5\right)H\bigr) \sigma-\trD\bigl(\gamma^\mu\left(1-\gamma_5\right)H\bigr) \sigma^\dagger\right]\!\lambda^{(b)}, 
  \end{equation}
where the right-handed current $j_R^{\mu,b}$ can be found from the left-handed current using parity
($H \to \gamma_0 H\gamma_0$,\ \  $\sigma\to\sigma^\dagger$).  We obtain the decay constant, or more 
precisely $\Phi\equiv f\sqrt{M}$, from the relation
\begin{equation}\label{eq:Phi0}
  \langle 0 \vert j_5^{\mu,a} \vert P_a\rangle =  -i  v^\mu  \Phi_a \qquad \textrm{(no sum on }a),
     \end{equation}
which implies $\Phi_a = \kappa$ to leading order when $\theta=0$.

When $\theta\not=0$, it affects $\Phi$ through the expectation value of $\sigma$ in \eq{HQaxial-current}:
$\langle\sigma\rangle = \sqrt{ \langle\Sigma\rangle} \not= I$.  Using \eq{vacuum} for $\langle\Sigma\rangle$
gives 
\begin{equation}\eqn{Phi-theta}
\Phi_x(\theta) = \Phi_x(\theta=0) \cos\left(\frac{\beta}{2}-\frac{5\delta}{4}\right),  
\end{equation}
for light valence quark $x$.   \Equations{betap}{deltap} then imply
\begin{equation}\eqn{Phi-result}
\Phi''_x = -\Phi_x(0)\frac{m^2m_s^2}{(m+2m_s)^2}\;\frac{1}{4m^2_x}.
\end{equation}
This smoothly connects to the light quark result for $f_{xy}''$, \eq{fxy-nf3}, in the limit $m_y\to\infty$.  (Note 
that the factor of $\sqrt{M}$ difference between $\Phi$ and $f$ is not important here, since $M$ has small $\theta$ dependence.)

At leading order, the heavy-light meson mass is independent of $\theta$, and has been removed from
$ \cL_{\chi,{\rm HL}} $, as usual.  However $\theta$ dependence can enter through the (NLO) light-quark
mass contributions to the Lagrangian,
\begin{eqnarray}\eqn{LHLm}
  \cL_{\chi,{\rm HL},m}  &=&  2\lambda_1 B_0\,\sTr(\overline{H} H \cM^+ ) +2\lambda'_1 B_0\, \sTr(\overline{H} H)\sTr(\cM^+ ), \\
\cM^+ &\equiv& \frac{1}{2} (\sigma \cM_B^*\sigma +\sigma^\dagger \cM_B \sigma^\dagger) \eqn{Mplus},
\end{eqnarray}
where $\cM_B$ is the light-quark mass matrix given in  
\eq{MB}), $\lambda_1$ and $\lambda'_1$ are new LECs, and $B_0$ (often omitted in  definitions of 
$\lambda_1,\lambda'_1$) is the light-quark LEC from \eq{ChPT-L}.  The dependence of the heavy-light
meson mass $M$ on the light valence mass is proportional to $\lambda_1$, while the sea-quark mass dependence comes from $\lambda_1'$.  

Plugging in $\langle\sigma\rangle$ to \eqs{Mplus}{LHLm}, and adding on the heavy-light mass in the chiral limit, $M_0$, which has been omitted from \eq{LHL}, gives
\begin{eqnarray}
M_x(\theta) &=& M_0 + 2\lambda_1 B_0 m_x \cos\left(\beta-\frac{5}{2}\delta\right) +\nonumber\\
&& + 2\lambda_1' B_0 \Big(2m \cos(\alpha-\theta/3) + m_s \cos(2\alpha+\theta/3)\Big).\eqn{MHL-theta}	
\end{eqnarray}
From \eqsthree{alphap-again}{betap}{deltap}, we then obtain
\begin{equation}\eqn{MHL-result}
M''_x = -2B_0\lambda_1\frac{m^2m_s^2}{(m+2m_s)^2}\;\frac{1}{m_x}
-2B_0\lambda'_1\frac{mm_s}{m+2m_s}.
\end{equation}
Note that fractional changes in $M_x$ with topology will be quite small (except in the limit when $m_x\ll m$), 
because $M_x$ is dominated by the $M_0$ term, which is independent of $\theta$.  
As in the light-light partially quenched case, our results for both masses, \eq{MHL-result}, and
decay constants, \eq{Phi-result}, vanish when either sea-quark mass goes to zero, consistent with
the discussion in Appendix \ref{m=0}.

To apply \eq{MHL-result} to lattice data, we  need the LECs $\lambda_1$, $\lambda'_1$ and $B_0$.
From the flavor splittings of
B or D mesons, we can extract $\lambda_1 \approx 0.2\; (\textrm{GeV})^{-1}$ (see, for example,
\rcite{Bazavov:2011aa}). 
A more detailed analysis of the heavy-light lattice data from the Fermilab/MILC Collaboration \cite{javad2017}
gives $\lambda_1=0.232(2) \; (\textrm{GeV})^{-1}$, where the error is statistical only.  The same analysis
implies $\lambda'_1=0.042(4) \; (\textrm{GeV})^{-1}$; the error is again statistical.
The smallness of $\lambda'_1$ is not surprising, since it is suppressed by large-$N_c$ counting.  
Because in addition 
the $\lambda'_1$ term does not blow up as $m_x\to0$, unlike the $\lambda_1$ term, its effects are negligible
at the currently available statistical precision. 
To obtain  $B_0$ we can use, for example,  $M_{xy}^2=B_0(m_x+m_y)$, where $M_{xy}$ is the mass of the light 
pseudoscalar meson made of $\bar x$ and $y$. 

The result for the 
mass in \eq{MHL-result} does {\em not}\/ join smoothly onto the corresponding light-light formula,
\eq{mxy-nf3}.  As $m_y$ gets heavy, $M''_{xy}$ falls like $1/\sqrt{m_y}$, where we have used $M_{xy}(0)=
\sqrt{B_0(m_x+m_y)}$.  However the heavy-light $M''_x$ is independent of the heavy-quark mass.  The 
difference can be traced to the simple fact that light-light and heavy-light meson masses have different
dependence on the masses of their valence quarks.  It is still true, though, that in both cases  $M''/M$ vanishes in the limit of infinitely heavy quark mass.  
 
\section{Staggered corrections}\label{sec:staggered}
 It is not difficult to include the leading discretization corrections from taste violations with rooted
 staggered quarks.  Each flavor becomes a staggered field with 4 tastes, and sea quarks are also replicated 
 $n_r$ times.  Rooting is accomplished by taking $n_r\to1/4$ at the end of the calculation 
 \cite{rooting-replicas,Bernard:2007ma}.
 
 We assume that the exact shift symmetry of staggered quarks  \cite{shift} does not get spontaneously broken
 when $\theta$ becomes nonzero.
 At the level of the chiral theory, shift symmetry corresponds to the discrete taste symmetry \cite{Bernard:2007ma}
 \begin{equation}\eqn{discrete-taste}
 \Sigma \to \xi_\mu \Sigma \xi_\mu \qquad \textrm{(no sum on $\mu$)},
 \end{equation}
 where $\xi_\mu$ ($\mu=1,\cdots,4$) is any of the generators of the taste algebra.  This symmetry is enough
 to guarantee that $\langle\Sigma\rangle$ has trivial dependence on taste:
  \begin{equation}\eqn{vac-sigma-taste}
  \langle\Sigma\rangle= \langle\Sigma\rangle_R\otimes \xi_I, 
\end{equation}
 where $\langle\Sigma\rangle_R$ is a ``reduced'' diagonal matrix in flavor and replica space only, and $\xi_I$ is the $4\times4$
 identity matrix in taste space.   
 
 The determination of  $\langle\Sigma\rangle_R$ in the 2+1 flavor case
 is then is very similar
 to the calculation of $\langle\Sigma\rangle$ in \secref{PQ}. In analogy with \eq{vacuum-general} we parameterize
 $\langle\Sigma\rangle_R$ by
 \begin{eqnarray}\eqn{vacuum-staggered}
\langle\Sigma\rangle_R &=&\exp\Big(i\,{\rm diag}(\alpha+\delta/n_r+i\epsilon/n_r,\cdots,\alpha+\delta/n_r+i\epsilon/n_r,
\cdots,
-2\alpha+\delta/n_r+i\epsilon/n_r, \cdots,\nonumber\\
&&\hspace{20mm}\beta-3\delta/2+i\epsilon,\;-\beta-3\delta/2+i\epsilon,\;
 5i\epsilon/2-i\gamma,\;5i\epsilon/2+i\gamma)\Big),
\end{eqnarray}
where $\cdots$ stands for the replication of the preceding entry $n_r-1$ times, and the explicit factors of
$1/n_r$ compared with \eq{vacuum-general} are necessary here for the stracelesness of the exponent.
 
 Aside from the  replication of the sea quark
 flavors, the main difference with the continuum calculation
 is the presence of the taste-violating contribution to the potential, $a^2 \cV$ \cite{LEE-SHARPE,AUBIN-BERNARD}.  
 Dependence on $\theta$  arises in some terms in $a^2 \cV$  both explicitly, though the anomalous chiral rotation that
 removes the $\theta F\! \tilde F$ term (see Appendix \ref{m=0}), and implicitly, though the expectation value
 of $\Sigma$.  However, because of the simple taste structure of  $\langle\Sigma\rangle$, the contributions
of $a^2 \cV$ combine with those of the quark mass term and produce terms proportional to
squared taste-singlet meson masses. In direct correspondence with  \eqsthru{alphap-again}{deltap}, we find
\begin{eqnarray}
\eqn{alphap-stag}
\alpha' &=& \frac{ M^2_{\pi, I}-M^2_{S, I} }{12n_r(M^2_{\pi, I}+2M^2_{S, I})} \ ,\\
\eqn{betap-stag}
\beta' &=& \frac{M^2_{X, I}-M^2_{Y, I}}{8n_r\,M^2_{X, I}M^2_{Y, I}}\cdot\frac{M^2_{\pi, I} M^2_{S, I}}{M^2_{\pi, I}+2M^2_{S, I}}=\hat\gamma'\equiv-i\gamma', \\
\eqn{deltap-stag}
\delta' &=& \frac{M^2_{X, I}+M^2_{Y, I}}{20n_r\,M^2_{X, I}M^2_{Y, I}}\cdot\frac{M^2_{\pi, I} M^2_{S, I}}{M^2_{\pi, I}+2M^2_{S, I}}=\hat\epsilon'\equiv-i\epsilon'\ ,
\end{eqnarray}
 where the taste-singlet meson masses are given by
 \begin{eqnarray}
 M^2_{\pi, I} &=& 2B_0 m + a^2 \Delta_I, \eqn{MpiI}\\
 M^2_{S, I} &=& 2B_0 m_s + a^2 \Delta_I, \eqn{MSI}\\
 M^2_{X, I} &=& 2B_0 m_x + a^2 \Delta_I,\eqn{MXI}\\
 M^2_{Y, I} &=& 2B_0 m_y + a^2 \Delta_I \eqn{MYI}.
\end{eqnarray}
 Here $a^2\Delta_I$ is the splitting of the taste-singlet mesons from the corresponding pseudo-Goldstone 
 (taste-$\xi_5$) mesons.   Note that
 \eqsthru{alphap-stag}{deltap-stag} reduce to \eqsthru{alphap-again}{deltap}
 after  taking the continuum limit and the rooting limit ($n_r\to1/4$). 
 
 We may then find the $\theta$ dependence of the valence $xy$ meson mass and decay constant
 by following at tree level the staggered $\theta=0$ calculations of  
 \rcites{Aubin:2003uc,AUBIN-BERNARD}.  We choose taste $\xi_5$ for the meson to correspond with the
 choice made  in the simulations.  We obtain, after rooting,
 \begin{eqnarray}
M_{xy,5}'' &=& -\frac{1}{4M_{xy,5}}
\frac{M^4_{\pi, I} M^4_{S, I}}{(M^2_{\pi, I}+2M^2_{S, I})^2}\;\left[\frac{M^2_{X,5}}{M^4_{X, I}} +\frac{M^2_{Y,5}}{M^4_{Y, I}}\right],
\eqn{mxy-stag}\\
f_{xy,5}'' &=&   -f_{xy,5}\; \frac{M^4_{\pi, I} M^4_{S, I}}{(M^2_{\pi, I}+2M^2_{S, I})^2}\;\frac{(M^2_{X, I}-M^2_{Y, I})^2}{4M^4_{X, I} M^4_{Y, I}}\ ,
\eqn{fxy-stag}
\end{eqnarray}
where all quantities on the right-hand sides are evaluated at $\theta=0$.  The subscript 5 indicates taste $\xi_5$.
The masses of these pseudo-Goldstone mesons are
\begin{eqnarray}
 M^2_{xy, 5} &=& B_0(m_x+m_y), \\
  M^2_{X, 5} &=& 2B_0 m_x ,\\
 M^2_{Y, 5} &=& 2B_0 m_y .
\end{eqnarray}
Note that the partially quenched singularities as $m_x$ or $m_y$ go to zero are now cut off at 
nonzero lattice spacing by the taste-singlet splitting.

Paralleling what occurs for the vacuum angles, the  $f_{xy,5}''$ result 
corresponds precisely to the continuum result, \eq{fxy-nf3}, with the simple replacement of each 
quark mass by the squared mass of the associated taste-singlet meson.   The same simple correspondence
between the leading order continuum and staggered results also occurs for the topological susceptibility
\cite{Billeter:2004wx}.  For the meson mass, however, the direct correspondence would occur only 
for the $\theta$ dependence of the taste-singlet mass.  The taste-$\xi_5$ squared mass gets an
explicit factor of each valence quark mass, which appear without the singlet splitting $a^2\Delta_I$, thereby 
producing the
$M^2_{X,5}$ and $M^2_{Y,5}$ terms in \eq{mxy-stag}.  The $M_{xy,5}$ term in the denominator 
arises simply from the fact that we give $M''_{xy,5}$ and not $(M^2_{xy,5})''$.

It is straightforward to extend these calculations to heavy-light systems.  In the heavy-light chiral
Lagrangian, staggered discretization effects appear only at NLO \cite{Bernard:2013qwa}.  This is contrast 
to the light-light Lagrangian, where the taste-violating
potential $a^2\cV$ is LO in the usual power counting $m_q \sim a^2$, where $m_q$ is a generic light-quark
mass.    For $\Phi$, whose $\theta$-dependence starts at LO, this means that the result
for $\Phi_x(\theta)$ in terms of the angles $\beta$ and $\delta$, \eq{Phi-theta}, remains valid.  We just
must use the staggered values for $\beta'$ and $\delta'$, \eqs{betap-stag}{deltap-stag}, to find $\Phi''_x$.
We obtain
\begin{equation}\eqn{Phi-stag}
\Phi_{x,5}'' =   -\Phi_{x,5}\; \frac{M^4_{\pi, I} M^4_{S, I}}{(M^2_{\pi, I}+2M^2_{S, I})^2}\;\frac{1}{4M^4_{X, I}}\ ,
\end{equation} 
where the subscript 5 indicates a taste-$\xi_5$ meson.

The calculation is a bit more complicated for the heavy-light meson mass because $\theta$-dependence 
first appears 
 at NLO.  At this order, there are also a large number of $a^2$ terms in the heavy-light chiral
Lagrangian, which are catalogued in \rcite{Bernard:2013qwa}, and appear
in the terms $\cL^{A1}_{2,a^2}$, $\cL^{B1}_{2,a^2}$, $\cL^{A2}_{2,a^2}$, and $\cL^{B2}_{2,a^2}$ defined there.
Although the majority of these terms
do not contribute to  $\theta$-dependence, there are ten terms that do, both explicitly and implicitly, as in
the light-light potential $a^2\cV$ discussed above.    Unlike what happens in 
the light-light case, however, the $a^2$ terms do not combine
with the quark-mass terms to form  taste-singlet light-light meson masses, because the
heavy-light  LECs are independent of the light-light ones.  We find
\begin{eqnarray}
M_{x,5}'' &=&   -\lambda_1\; \frac{M^4_{\pi, I} M^4_{S, I}}{(M^2_{\pi, I}+2M^2_{S, I})^2}\;\frac{(M^2_{X,5}+a^2\Delta^{\rm val}_{\rm HL})}{M^4_{X, I}} +\nonumber \\
&& -\lambda'_1\; \frac{2(M^2_{\pi,5}+a^2\Delta^{\rm sea}_{\rm HL})M^4_{S, I}+
 (M^2_{S,5}+a^2\Delta^{\rm sea}_{\rm HL})M^4_{\pi, I}}{(M^2_{\pi, I}+2M^2_{S, I})^2}
 \eqn{MHL-stag}
\end{eqnarray} 
Here $\Delta^{\rm val}_{\rm HL}$ is a linear combination of the LECs  $K^{A1}_{1,3}$, $K^{A1}_{1,4}$, $K^{A2}_{1,2}$, $K^{A2}_{1,3}$, $K^{A2}_{1,7}$, $K^{A2}_{1,8}$, $K^{B2}_{1,1}$, and $K^{B2}_{1,2}$ from 
\rcite{Bernard:2013qwa} (divided by $\lambda_1$),
and $\Delta^{\rm sea}_{\rm HL}$ is a linear combination of the LECs  $K^{A1}_{2,3}$ and $K^{A1}_{2,4}$
(divided by $\lambda'_1$).  We have not bothered to work out the coefficients in these linear combinations
since the relations are unlikely to be useful, but it is straightforward to find them if they are ever needed.
One can easily check that \eqs{Phi-stag}{MHL-stag} reduce to \eqs{Phi-result}{MHL-result}, respectively, in the
continuum limit.

There are also other, ``generic,'' discretization effects with staggered quarks that have nothing to do
with the (partial) violation of chiral symmetry that results in taste splittings.  Such generic effects are
of order $\alpha_s a^2$ in a tree-level improved staggered action; the two-stage smearing in the HISQ action 
further suppresses these effects by a numerical factor.
Analyses of various physical quantities in HISQ simulations
typically give sub-percent generic discretization errors for the
range of lattice spacings ($a\ltwid 0.09$ fm) considered here.  For example, $f_K/f_\pi$ varies
from its continuum limit  by about 0.3\% over these lattice spacings \cite{fermimilcdecay2014, Bazavov:2017lyh}.

\section{Comparison to simulation results} \label{sec:compare}

The calculation of meson masses and decay constants on the HISQ ensembles is described
in Ref.~\cite{fermimilcdecay2014}.
To find the dependence on the
topological charge, we use the results of a single-elimination jackknife analysis of these
quantities together with the time histories of topological charge shown above.
To estimate $B''$ we rearrange Eq.~\ref{eq:topo_dep}, using $<Q^2>=\chi_t V$, as
\BNE \eqn{topo_dep2}
B \big|_{Q,V} = \LP B + \frac{1}{2<Q^2>} B'' \RP \,-\, \LP \frac{1}{2 <Q^2>^2} \,B'' \RP\,Q^2 , \ENE
This lets us find $B''$ by fitting $B \big|_{Q,V}$ to a constant plus quadratic in $Q$.
In fact what is available is a list of single elimination jackknife averages for $B$ ---
that is, values for $B$ obtained by omitting one lattice from the analysis.
We can effectively ``undo'' the jackknife using
\BNE \eqn{jack_reverse} B_j - \overline B = - \LP N-1 \RP \LP \overline B_j - \overline B \RP \ENE
where $\overline B = \frac{1}{N} \sum_i B_i$ is the full sample average and
$\overline B_j = \frac{1}{N-1} \sum_{i \ne j} B_i$ is the $j$'th jackknife sample.
In practice we simply fit the jackknife averages, and supply the factor of $-(N-1)$
later.  To estimate the error on $B''$ we assign an error equal to the error in
our sample average $\overline B$ to each data point, which results in a fit with
$\chi^2 \approx N$, and then use the error on the fit parameters found in the
standard way.  (This actually neglects the part of the variance of $B$ coming from
its dependence on $Q^2$, but in practice this turns out to be a small part of
the variance.)

Although the results are barely statistically significant, they are consistent with the \chpt\ predictions.  Statistically
significant signals are found in the $m_l=m_s/5$ ensembles, since these have much smaller
physical volumes than the physical light quark mass ensembles.
The effects that we observe, in \eq{topo_dep}, 
all have an overall factor of $1/V$, even if the parenthesized factor $\LP 1 - \frac{Q^2}{\chi_T V} \RP$ covers a range of order one.
For example, in the $a\approx 0.06$ fm
ensembles the $m_l=m_s/5$ lattices have a volume of $180$ (fm)$\null^4$, while the physical
$m_l$ lattices have a volume of $1920$ (fm)$\null^4$.  Also, \eq{fxy-nf3} shows
that the derivatives of the masses and decay constants have a partially quenched divergence
when $m_x$ or $m_y$ goes to zero with $m_l$ fixed, and for the $m_l=m_s/5$ ensembles we
have used valence quark masses smaller than $m_l$, in some cases as small as the
physical $m_l$.

\begin{figure}
\vspace{-1.0in}
\begin{center} \includegraphics[width=.57\textwidth]{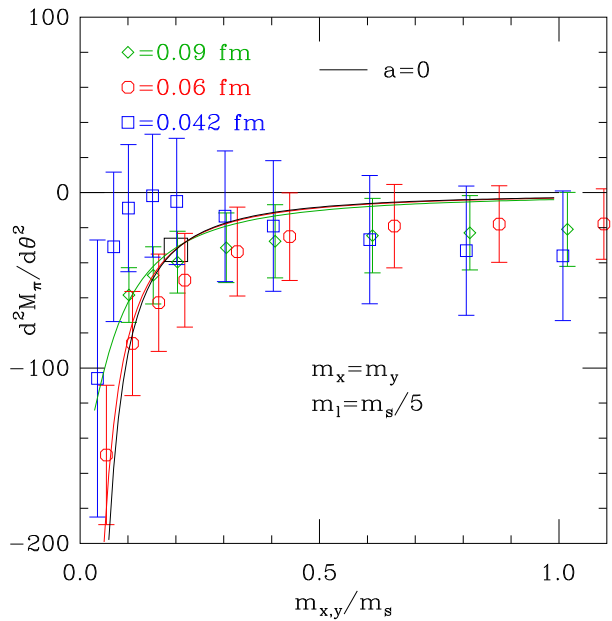} 
\end{center}
\vspace{-1.05in}
\caption{
\label{d2m_fig}
$\PARTWO{M}{\theta}$ on ensembles with $m_l=m_s/5$
as a function of  $m_x$, with $m_y=m=m_x$.
The black line is the \pqchpt\  prediction at $a=0$, or without taste breaking effects (no free parameters).
The green and red lines show the \pqchpt\ prediction including taste breaking, Eq.~\protect\ref{eq:mxy-stag}, for the
0.09 and 0.06 fm ensembles.
The black square marks the unitary point, with valence quark mass equal to the sea quark mass.
}
\end{figure}

Figure~\ref{d2m_fig} shows $\PARTWO{M}{\theta}$ for the $m_s/5$ ensembles for degenerate valence quark
masses, $m_x=m_y$.  The black line in the figure is the \pqchpt\ prediction in \eq{mxy-nf3},
which we emphasize is a prediction with no free parameters.
Obviously the statistical errors are large, but they are consistent with the prediction, and the
divergence at small valence quark mass is clearly seen.
The green and red lines in this figure show the \pqchpt\ prediction including the taste-breaking effects, Eq.~\ref{eq:mxy-stag}.
To calculate these effects we need to know the
 taste splitting $\Delta_I$ in \eqsthru{MpiI}{MYI}, which is expected to be proportional to $\alpha_s^2$.
 Since $\alpha_s$ changes significantly over this range of 
lattice spacings, we simply use the values of
 $\Delta_I$
computed directly from $M_I^2-M_5^2$ at $a=0.09$ fm and $a=0.06$ fm.  
We do not have a direct measurement of taste breaking on
the 0.042 fm ensemble, but we expect the result there to be very close to the continuum result, as is clear
from the fact that $a=0.06$ fm (red) curve is already barely distinguishable from the continuum (black) curve.

\begin{figure}
\vspace{-1.0in}
\begin{center}\begin{tabular}{ll}
\hspace{-0.4in} \includegraphics[width=.57\textwidth]{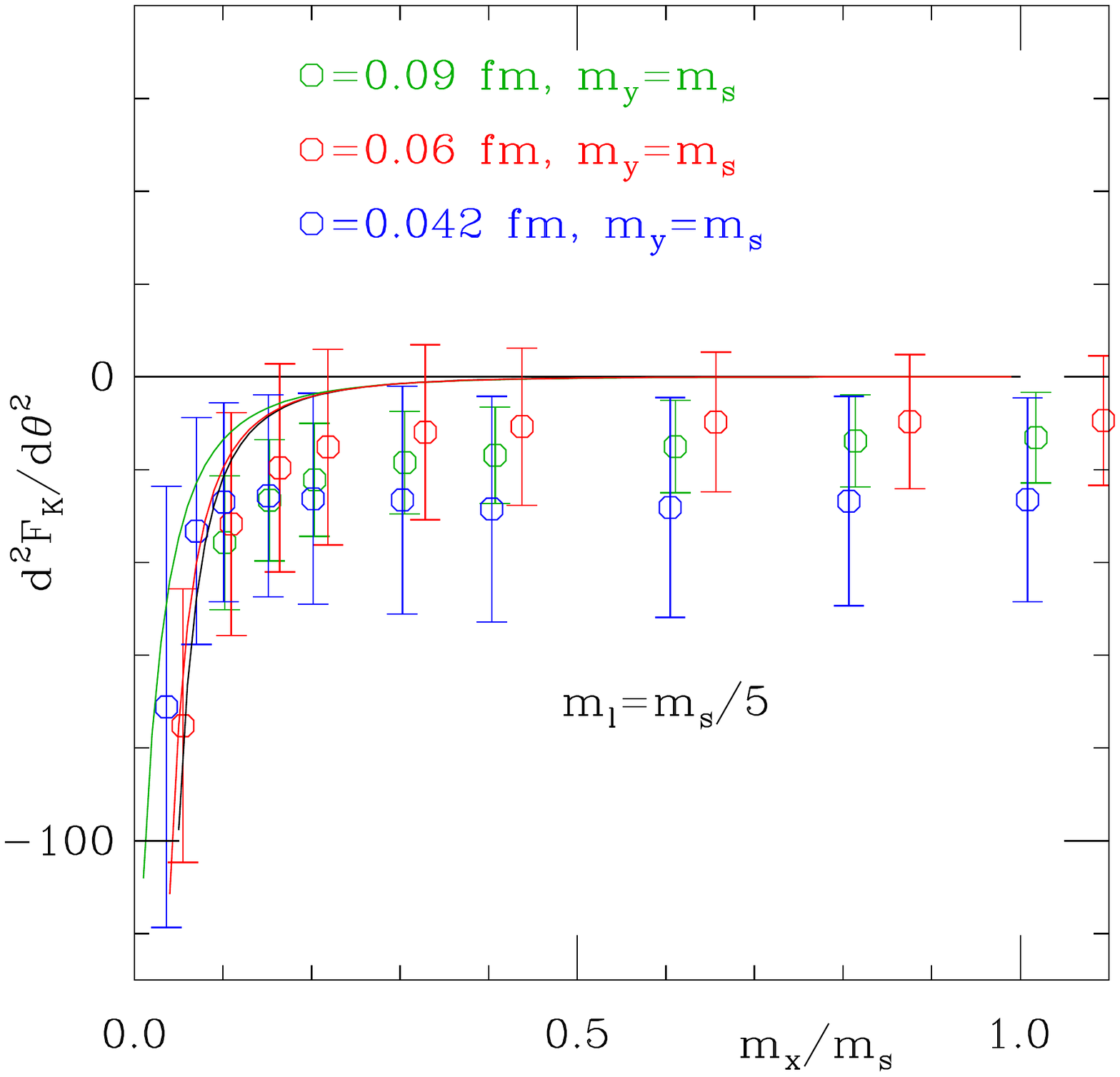} &
\hspace{-0.5in} \includegraphics[width=.57\textwidth]{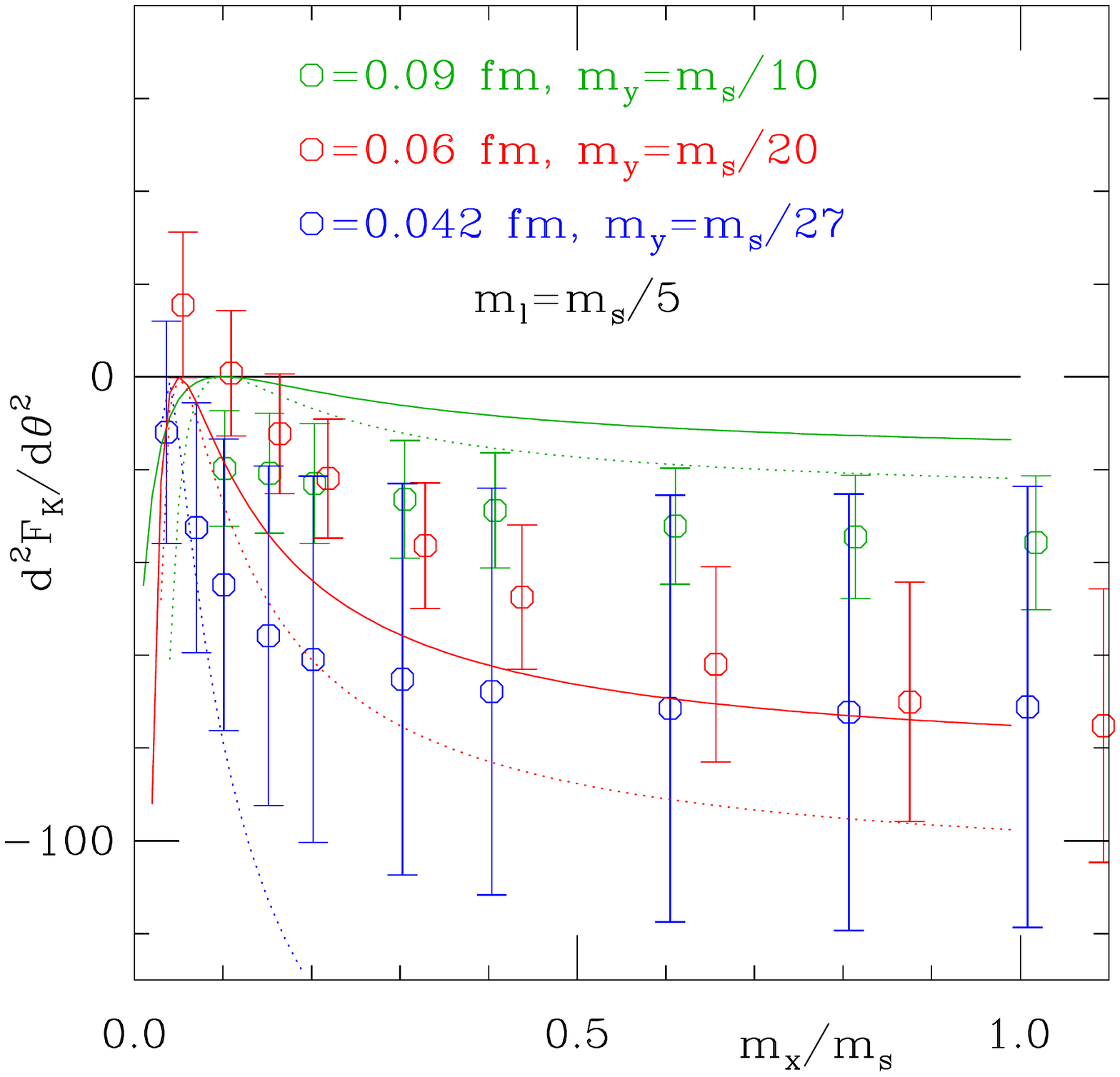} \\
\end{tabular}\end{center}
\vspace{-1.05in}
\caption{
\label{d2f_fig}
$\PARTWO{F}{\theta}$ on ensembles with $m_l=m_s/5$.
The left panel shows $\PARTWO{F}{\theta}$ as a function of one valence
quark mass, $m_x$ along the line $m_y=m_s$.
The black line is the \pqchpt\ prediction (no free parameters), which vanishes for degenerate quarks.
The red ($a=0.06$ fm) and green ($a=0.09$ fm) lines include the effects of taste breaking from Eq.~\protect\ref{eq:fxy-stag}.
The right panel shows the quantity  with $m_y$
fixed to the smallest available value.
The dotted lines are \pqchpt\ predictions ignoring taste breaking, or at $a=0$; there are three separate lines
because the smallest valence quark mass is different in each ensemble: 
$0.1\, m_s$, $0.05\,m_s$ and $0.037\,m_s$ for the $0.09$, $0.06$ and $0.042$ fm ensembles
respectively.
Again, the solid red and green lines include the effects of taste breaking from Eq.~\protect\ref{eq:fxy-stag}.
}
\end{figure}

Since $\PARTWO{F}{\theta}$ vanishes for degenerate valence quarks, we plot this quantity
along different lines in Fig.~\ref{d2f_fig}.  The left panel shows $\PARTWO{F}{\theta}$
as a function of one valence quark mass, $m_x$, with the other fixed at the strange quark
mass, together with the \chpt\ prediction.
The right panel shows $\PARTWO{F}{\theta}$
along lines where $m_y$ is held fixed at the lightest valence quark mass available in
each ensemble.  The vanishing of $\PARTWO{F}{\theta}$ when the valence quarks are
degenerate is particularly striking in this plot.
In the left panel, with $m_y=m_s$, the black line shows the continuum result, without taste
breaking, and the green and red lines the results for 0.09 and 0.06 fm including the taste
breaking.  In the right panel, since each of these three ensembles had a different lightest
valence mass, the colored dotted lines show the prediction without including taste breaking
for each of the three ensembles.  Because 
 one of the valence quark masses, $m_y$, is held fixed at
its lightest values in the right hand panel, the effects of taste violation are large for all values of $m_x$.  As
a fraction of the continuum value, the staggered effects
in this graph do not decrease as much as might be
expected when $a$ changes from 0.09 fm to 0.06 fm because the relative size of $a^2 \Delta_I$
and $2B_0 m_y$ in  $M_{Y,I}$ in \eqs{MYI}{fxy-stag}  is what matters, and
$m_y$ has changed from $m_s/10$ to $m_s/20$.

Our statistical errors on the heavy-light masses and decay constants are much larger than
on the light quark quantities, so we are unable to test our data against the chiral perturbation theory for the heavy-light quantities.

Knowing the dependence of masses and decay constants on the average $Q^2$, we can correct
our simulation results to account for the difference of the average in our simulation, $\langle Q^2\rangle _{sample}$, 
and the correct $\langle Q^2\rangle $.  To estimate this correct $\langle Q^2 \rangle$ we use the lowest order 
staggered \chpt\ result \cite{Billeter:2004wx}
\begin{equation}\eqn{chiT-stag}
\chi_T = \frac{f_\pi^2}{4(2/M_{\pi,I}^2 + 1/M_{S,I}^2)}
\end{equation} 
where the taste-singlet masses $M_{\pi,I}^2$ and $M_{S,I}^2$ are given in \eqs{MpiI}{MSI}.
The \chpt\ results are shown in Fig.~\ref{qsq_vs_a_fig}.
In computing the \chpt\ form for the ensembles with $a > 0.042$ fm, we used measured values for the taste singlet pion
and $s\bar s$ pseudoscalar masses on each ensemble.  For the $0.042$ fm physical quark-mass
ensemble we estimated taste breaking effects by scaling the taste splittings from the $0.06$ fm
physical quark-mass ensemble, and for the $0.042$ and $0.03$ fm, $m_s/5$ ensembles, where we expect the taste-breaking effect
to be negligible (and certainly not measurable with our statistics), we used the Goldstone
pseudoscalar masses.
For large $a$, the deviation from
the lowest-order \chpt\ result is due to lattice artifacts, probably mostly higher-order
taste-breaking effects, but for $a=0.042$ and $0.03$ fm we expect the \chpt\ results to be
pretty good. 

For examples of the size of this effect in our simulations, we look at the two ensembles with $a \approx 0.042$ fm.
For example, to adjust the decay constants, rearrange \eq{topo_dep} as
\BNE f_{corrected} = f_{sample} - \frac{1}{2\chi_T V} f^{\prime\prime} \LP 1 - \frac{\langle Q^2\rangle _{sample}}{\chi_T V} \RP.  
\eqn{correction}\ENE
Table~\ref{tab:toposhifts} shows the size of the topology adjustment for selected quantities, together with the central value and
statistical errors.  The sign of the adjustment differs between the two ensembles because, as can be seen in Fig.~\ref{qsq_vs_a_fig},
the difference between the sample average $Q^2$ and the chiral perturbation theory prediction is different in the two cases.
The effects are
larger in the ensemble with $m_l/m_s=0.2$, since these lattices have
much smaller volume and partial quenching divergences.
It is amusing to note that for the physical quark mass ensemble the topology adjustment for
$f_K/f_\pi$ is a factor of $\sim\!6$ smaller than 
the ``conventional'' finite size effects from pions propagating around the
periodic lattice, estimated in NLO staggered \chpt, of 0.0009. (Conventional finite size effects on the
heavy quark quantities are quite small.)

\begin{table} \begin{center}
\caption{ \label{tab:toposhifts}
Examples of topological adjustments for pseudoscalar masses and decay constants in the $0.042$ fm HISQ ensembles.
Each field contains the unshifted value for the quantity, its statistical error in parentheses, and
the topology adjustment in square brackets.  These quantities are evaluated at valence masses equal to the
sea quark masses, which have small tuning errors.  For the heavy-light masses we used $B_0 = 3.4$ GeV,
$\lambda_1 = 0.232$ (GeV)$\null^{-1}$ and $\lambda_1^\prime = 0.042$ (GeV)$\null^{-1}$ \cite{javad2017}.
Note that since the $Q^2$ in the second line is the $Q^2$ averaged over our sample, there is
no statistical error associated with it.
}
\begin{tabular}{|c|l|l|}
\hline
	& $m_l=m_s/5$ & $m_l=\mathrm{physical}$ \\
\hline
$Q^2_\mathrm{sample}/Q^2_{\chi\mathrm{PT}}$ & 1.30 & 0.65 \\
\hline
$f_K/f_\pi$ & 1.20508(0.00250)[-0.01271] & 1.19680(0.00114)[0.00015] \\
$aM_\pi$ & 0.031147(0.000172)[-0.000707] & 0.028964(0.000020)[0.000008] \\
$af_D$ & 0.048858(0.000261)[-0.000552] & 0.045389(0.000245)[0.000006] \\
$aM_D$ & 0.409786(0.000391)[-0.000044] & 0.400678(0.000258)[0.000001] \\
$af_{D_s}$ & 0.054828(0.000068)[-0.000001] & 0.053582(0.000025)[0.000000] \\
$aM_{D_s}$ & 0.430966(0.000116)[-0.000004] &  0.422041(0.000037)[0.000000] \\
\hline
\end{tabular}
\end{center} \end{table}

We note that this strategy is in the same spirit as our treatment of conventional
finite size effects.  We use \chpt\ to estimate the effects and correct our results, and estimates
of the effects of higher order \chpt\ and/or uncertainties in the \chpt\ parameters should be
included in the systematic error budget.

\section{Conclusions}
\label{sec:conclusions}

Our key \chpt\ results are given in \eqs{mxy-nf3}{fxy-nf3} for partially quenched light-light meson masses
and decay constants, and in \eqs{MHL-result}{Phi-result} for the corresponding quantities for heavy-light mesons.
In the light-light case, these results reproduce those of Aoki and Fukaya \cite{aokifukaya2009}, but are
now computed within a nonperturbatively-justified partially quenched chiral theory.  Results with the leading
staggered discretization effects included are given in \eqsfour{mxy-stag}{fxy-stag}{Phi-stag}{MHL-stag}.
The results can be compared with simulation data for the quantities as a function of 
topological charge $Q$ using \eq{topo_dep}.  We have done this in Figs.~\ref{d2m_fig} and \ref{d2f_fig};  there are
large statistical errors but the qualitative agreement is good.  Discretization effects are generally quite
small at the fine lattice spacings where these results are likely to be used. For very light valence-quark masses in
comparison to sea-quark masses, however, the staggered taste splitting cuts off the partially quenched singularity and can thus be numerically important.
  One can also  use  the results in conjunction with 
\eq{correction} to adjust quantities for incorrect sampling of $Q$, as discussed at the end of
\secref{compare}. The corrections in the cases we have looked at are very small 
(much less than the statistical errors) for our physical-mass ensemble at $a\approx 0.042$.
For the ensemble with $m_l=m_s/5$ at the same lattice spacing, however, the corrections are often
statistically significant, with the largest ($\sim\! 5$ times the statistical sigma) occurring for $f_K/f_\pi$, which has very small
statistical errors.   Both the smaller volume of this ensemble and the partial quenching divergences play
a role here.  In view of the fact that the corrections are calculated only to leading order in \chpt, one should
take a relatively large fraction of the correction as the associated systematic error.  Nevertheless,
our analysis suggests that it will be possible to satisfactorily control the
systematic errors due to non-equilibrated topological charge distributions, even at rather small lattice spacings.

\vspace{-0.2in}
\section*{Acknowledgements}
\vspace{-0.1in}

Computations for this work were carried out with resources provided by
the USQCD Collaboration,
the National Energy Research Scientific Computing Center,
the Argonne Leadership Computing Facility, 
the Blue Waters sustained-petascale computing project,
the National Institute for Computational Science,
the National Center for Atmospheric Research,
the Texas Advanced Computing Center,
and Big Red II+ at Indiana University.
USQCD resources are acquired and operated through funding from the Office of Science of the U.S. Department of Energy.
%
The National Energy Research Scientific Computing Center is a DOE Office of Science User Facility supported by the
Office of Science of the U.S. Department of Energy under Contract No.\ DE-AC02-05CH11231.
%
An award of computer time was provided by the Innovative and Novel Computational Impact on Theory and Experiment (INCITE)
program. This research used resources of the Argonne Leadership Computing Facility, which is a DOE Office of Science
User Facility supported under Contract DE-AC02-06CH11357.
%
The Blue Waters sustained-petascale computing project is supported by the National Science Foundation (awards OCI-0725070 and
ACI-1238993) and the State of Illinois.
Blue Waters is a joint effort of the University of Illinois at Urbana-Champaign and its National Center for Supercomputing
Applications.
This work is also part of the ``Lattice QCD on Blue Waters'' and ``High Energy Physics on Blue Waters'' PRAC allocations supported
by the National Science Foundation (award numbers 0832315 and 1615006).
This work used the Extreme Science and Engineering Discovery Environment (XSEDE), which is supported by National Science Foundation
grant number ACI-1548562~\cite{XSEDE_REF}.
Allocations under the Teragrid and XSEDE programs included resources at the National Institute for Computational Sciences (NICS) at
the Oak Ridge National Laboratory Computer Center, The Texas Advanced Computing Center and the National Center for Atmospheric
Research, all under NSF teragrid allocation TG-MCA93S002.
Computer time at the National Center for Atmospheric Research 
was provided by NSF MRI Grant CNS-0421498, NSF MRI Grant CNS-0420873, NSF MRI Grant CNS-0420985, NSF sponsorship of the National
Center for Atmospheric Research, the University of Colorado, and a grant from the IBM Shared University Research (SUR) program.
Computing at Indiana University is supported by Lilly Endowment, Inc., through its support for the Indiana University Pervasive
Technology Institute.

We wish to thank Maarten Golterman, Andreas Kronfeld,
and our colleagues in the MILC Collaboration for
helpful discussions, and, in addition, our MILC colleagues for developing the computer codes used in the project, and for generation
of the lattice ensembles used here.  We also thank Javad Komijani for pointing out an error in an earlier version of the heavy-light
analysis.  We are grateful to Maarten Golterman for a
critical reading of this manuscript and many helpful suggestions for its improvement.  Finally, we thank
the referee for suggesting we include a study of the staggered discretization corrections.

\appendix
\section{Vanishing sea-quark mass}
\label{m=0}

The usual expectation is that all $\theta$-dependence should disappear when one (or more) sea-quark 
masses vanish.  Indeed that is the reason that a zero value for the up-quark mass would solve the
strong CP problem \cite{strongCP}.  However, we will see in this appendix that the absence of
$\theta$-dependence is in general true only for spectral quantities in QCD, \ie quantities that are entirely
determined by the QCD Lagrangian.  While the second derivatives with respect to $\theta$ should vanish for all 
meson masses in the limit of a vanishing sea-quark mass, this is not necessarily true for decay constants, which 
depend on external (axial) currents.  In particular, the  results for decay constants in unitary theories,
\eqs{fpipp}{fK-nf3}, do not vanish when one sea-quark 
mass goes to zero.  On the other hand, as long as the valence masses
remain nonzero, derivatives of the partially-quenched decay constant, \eq{fxy-nf3}, do vanish in this limit.

To understand what is going on, it is helpful to look in detail at the chiral transformations that have (implicitly) 
been used to put the Lagrangian in various convenient forms.  
We work here in the partially-quenched context so that the results will apply to all the calculations 
described above.  Under a chiral transformation,
\begin{equation}\eqn{chiral-sigma}
\Sigma \to L\Sigma R^{-1},\qquad \Sigma^{-1} \to R\Sigma^{-1} L^{-1}.
\end{equation}
The axial current transforms as
 \begin{equation}
\cA_\mu^{ij} = i\frac{f^2}{4}\!\left(\partial_\mu\Sigma\Sigma^{-1}\! +
\Sigma^{-1} \partial_\mu\Sigma\right)^{ij} 
\to\; i\frac{f^2}{4}\!\left(L\partial_\mu\Sigma\Sigma^{-1}L^{-1} \!+
R\Sigma^{-1} \partial_\mu\Sigma R^{-1} \right)^{ij}. \eqn{PQaxial-current}
\end{equation}
In the partially-quenched Lagrangian, the chiral transformation is equivalent to leaving $\Sigma$
unchanged and transforming the mass matrix as:
\begin{equation}\eqn{chiral-M}
\cM \to L^{-1}\cM R,\qquad  \cM^* \to R^{-1}\cM^* L.
 \end{equation}
 Note that this is the inverse of  the fake (spurion) transformation on $\cM$ that would leave the 
 Lagrangian invariant.  

The first chiral transformation we consider (call it ``A'') is the anomalous, flavor-singlet one that removes the
 $\theta F\tilde F$ term from the QCD Lagrangian, and puts a uniform phase in the mass matrix, as
 in \eq{ChPT-L}.
 \begin{equation} \eqn{transA}
R_A = L^{-1}_A = \exp(\frac{i\theta}{6}I),\qquad \cM \to \cM_A \equiv \exp(\frac{i\theta}{3})\cM.
\end{equation}
where $I$ is the identity matrix given in \eq{T7}, and we have specialized to $N=3$ and $N_v=2$.  
All axial currents are invariant under this flavor-singlet transformation.

The second chiral transformation (``B''), is the one that removes the $\theta$-dependence 
from the valence- and ghost-quark masses.  This is the non-anomalous transformation
\begin{equation} \eqn{transB}
R_B = L^{-1}_B = \exp(\frac{-i\theta}{6}t),
\end{equation}
with $t$ given in \eq{t}. Transformation B produces the mass matrix $\cM_B$ given in \eq{MB} and used in the calculations 
of \secref{PQ}.  It leaves valence-valence currents (or sea-sea currents) unchanged,
so $f''_{xy}$ in \eq{fxy-nf3} is correct for the axial current defined as usual, with no $\theta$-dependence.

It is straightforward to check that we get the same answer for $f''_{xy}$ using mass matrix
$\cM_A$, \eq{transA}, instead of $\cM_B$.  As a further check, we may calculate the $\theta$-dependence
of the (strange) sea-valence decay constant, $f''_{sx}$.  In this case it is crucial to include the
nontrivial $\theta$-dependence of the axial current induced by transformation B {\it via}\/ \eq{PQaxial-current}.  Once this is done, the results in the A and B cases agree, and agree with the result for $f''_{xy}$,
\eq{fxy-nf3}, when
$m_x=m_s$. The meson
mass is of course insensitive to the currents, so there are no subtleties in the calculation, and
the results in cases A and B are again identical.  

We can now turn to the question of $\theta$-dependence when a sea-quark mass vanishes.  For
 definiteness in our $N=3$ example, let us take $m_s\to 0$.  The trick here is to make a third chiral transformation
 (``C'') that is non-anomalous and puts all $\theta$-dependence into the $m_s$ term in the quark mass 
 matrix:
 \begin{eqnarray} \eqn{transC}
R_C = L^{-1}_C &=& \exp(\frac{-i\theta}{6}\lambda),\qquad
 [\lambda \equiv {\rm diag}(1,1,-2,0,0,0,0)],\\
 \cM_B &\to& \cM_C = {\rm diag}(m, m, e^{i\theta}m_s, m_x, m_y, m_x, m_y).  \eqn{MC}
\end{eqnarray}
Now all $\theta$-dependence disappears for spectral quantities in the limit $m_s\to0$.  The valence-valence
meson mass provides an example:  $M''_{xy}$ in \eq{mxy-nf3} vanishes in this limit, for any
(fixed, nonzero) values of the other masses.  It is important not to take a valence mass
to zero before $m_s\to0$; the limits are not interchangeable because of the PQ singularities.  We can,
however, put $m_x=m_s,\ m_y=m$ or $m_x=m_y=m$ to get the full theory $M_K''$ or $M_\pi''$, 
respectively, \eqs{mK-nf3}{mpi-nf3}, which again vanish in the $m_s\to0$ limit.

Transformation C does not affect 
the valence-valence axial current, so $f''_{xy}$, \eq{fxy-nf3}, also vanishes as $m_s\to 0$.  However,
this is not true of the full-theory $f''_K$, \eq{fK-nf3}.  In this case, the axial current $\cA^{13}_\mu$
is changed by the transformation. Indeed, all $\theta$-dependence in the limit $m_s\to0$ comes from the 
current, and we find $f_K(\theta) = f \cos(\theta/2)$.  This gives the nonvanishing result $f''_K = -f_K(0)/4$
in this limit, in agreement with  \eq{fK-nf3}.  

We can similarly check the $m\to0$ limit.  In this case, we should put the $\theta$-dependence equally
into the up- and down-quark entries of the mass matrix, so as not to spoil isospin invariance, which 
was assumed in the calculations of \secref{PQ}.  We then find $f_K(\theta) = f \cos(\theta/4)$ in this limit,
giving $f''_K = -f_K(0)/16$, in agreement with the limit of \eq{fK-nf3}.  Further, $M_K''$ and $M_\pi''$
should vanish in this limit, in agreement with \eqs{mK-nf3}{mpi-nf3}.  (For $M_\pi''$, we need to use the
fact that $M_\pi$ itself vanishes in the limit.)

Finally, a note of warning:
The various limits of vanishing quark mass are subtle, and it is easy to go astray.  This is already clear
in the full theory from the fact that the limits $m\to 0$ and $m_s\to 0$ do not commute for 
$f''_K$, \eq{fK-nf3}.   Another interesting example is the limit $m_x\to0$ for a valence mass,
with sea masses and other valence masses held fixed.  We
expect a partially-quenched singularity in this case, and \eq{mxy-nf3} shows this for the spectral
quantity $M''_{xy}$. However, we can also make a plausible-sounding argument that $M''_{xy}$ should
vanish in this limit!  Starting from case B, suppose we make a non-anomalous transformation (``D'') 
\begin{eqnarray} \eqn{transD}
R_D = L^{-1}_D &=& \exp(\frac{-i\theta}{6}\tilde\lambda),\qquad
 [\tilde\lambda \equiv {\rm diag}(1,1,1,-3,0,0,0)],\\
 \cM_B &\to& \cM_D = {\rm diag}(m, m, m_s, e^{i\theta}m_x, m_y, m_x, m_y).  \eqn{MF}
\end{eqnarray}
Now all $\theta$-dependence, for spectral quantities, is in the valence-quark $x$ term in $\cM_D$,
so shouldn't $M''_{xy}$ vanish in the $m_x\to 0$ limit?   

The  problem with this argument is that the partially-quenched singularity is so strong as $m_x\to0$
that it overwhelms the reduction in spectral quantities coming directly from the  $e^{i\theta}m_x$ term in the
quark mass matrix. 
Keeping $m_x\not=0$ and repeating  the 
calculational steps in \secref{PQ}, we in fact reproduce \eq{mxy-nf3} for $M''_{xy}$.  
This is not unexpected because
we have simply made a non-anomalous chiral transformation, which should not affect physical quantities. 
Indeed, if we are careful about the phase introduced in the axial current by transformation D, we
also reproduce $f_{xy}''$ in \eq{fxy-nf3}. Further, despite the apparent breaking of quark-ghost symmetry
by   $\cM_D$, it is still true that $\langle\Sigma\rangle_{xx}= \langle\Sigma\rangle_{\tilde x\tilde x}$, just as
in the cases where we preserve  quark-ghost symmetry explicitly in the mass matrix.

One may still wonder whether we can accept that there is a 
discontinuity at $m_x=0$
and simply set $m_x=0$ from the start.   This is not allowed, however, because the ghost integral needs
a non-zero mass term for convergence.%
\footnote{We thank Maarten Golterman for this point.}

\section{Convergence of the neutral ghost-antighost integrals}
\label{ghost-mass}

The bosonic path integrals over the field $\tilde\phi$ for ghost-antighost mesons, \eq{Phip},
and the field  $\epsilon$ for the ``ghost-like'' neutral meson, \eqs{Phi}{T6}, must be convergent
in order for PQ\chpt\ to be nonperturbatively well-defined. Further,
in order for the perturbative vacuum defined by $\tilde\phi=0$, $\epsilon=0$ to be the correct one, 
it seems that 
$\Re(V)$ should have a minimum at this point for real $\epsilon$ and Hermitian $\tilde\phi$.  These conditions 
apparently require that the real part of the Lagrangian, expanded to quadratic order in
the fields, should be positive definite.   Indeed, the factors of
$i$ in \eqs{Phi}{Phip} are inserted to ensure that $\tilde\phi$ and $\epsilon$ have positive
kinetic energy terms for $p^2 \not=0$.  The same factor of $i$ also guaranties that the real part of the mass
term of $\tilde\phi$ is positive definite as long as no valence masses vanish.

The ghost-like field $\epsilon$, however, presents problems. Because it is a linear combination of
quark-antiquark and ghost-antighost fields,   there is a competition in its mass term between  ghost
masses, which give positive terms, and quark masses, which give negative terms. Each ghost
term always wins over the corresponding valence term, because $\epsilon$ has more support in the
ghost sector than in the valence sector.  In contrast, the competition with the sea masses can go
either way.   For low enough
valence and ghost masses compared to sea masses, the real part of the $\epsilon$ mass term
will become negative, putting into doubt the convergence of the $\epsilon$ path integral, or at least
the validity of the perturbative vacuum at $\epsilon=0$.  This leads to the Sharpe-Shoresh lower bound \cite{sharpeshoresh2001} on the  valence masses.  Because $\epsilon$ can mix with the neutral component of
$\tilde \phi$, corresponding for example to the parameter $\gamma$ in \eq{vacuum-general}, 
\rcite{sharpeshoresh2001} requires that the full neutral ghost and ghost-like mass matrix be positive definite, 
resulting in the bound
\begin{equation} \eqn{SSbound}
N_v\overline{\chi^{-1}_v}(N_v\overline{\chi_v} + N\overline{\chi}) < (N+N_v)^2,
\end{equation}
 where $\overline{\chi}$ and $\overline{\chi_v}$ are the average sea-quark mass and average valence-quark
 mass, respectively, and $\overline{\chi^{-1}_v}$ is the average inverse valence-quark mass.  
 Sharpe and Shoresh suspect that this is some kind of artifact of the chiral theory,
 since the underlying partially quenched QCD has no apparent problem at
 or below this bound. Nor is there any  evidence from the calculation of standard 
 perturbative quantities within partially quenched \chpt, or their comparison with simulations, that things
 go wrong when the bound is violated.   Nevertheless, since much of the simulation
 data that we analyze violates this bound, the apparent lack of convergence of the chiral theory
 is disconcerting.  We certainly cannot claim our chiral results to be nonperturbatively
 correct in the region where the bound is violated unless we can show that the bound itself is not actually
 an obstacle to using the theory around the standard vacuum.
 
 We work primarily in the case $\theta=0$; nonzero but small $\theta$ does not present any significant additional
  problems.  
 We also start by considering a simpler theory than that of \secref{PQ}, with $N=2$, $N_v=1$ and degenerate 
 sea-quark masses.  After showing  
 in this simple model that violation of the Sharpe-Shoresh bound does not lead to any 
 problem with convergence or with the perturbative vacuum, we will be able to use a shortcut to arrive at a similar 
 conclusion for the case of interest, $N=3$ and $N_v=2$ with nondegenerate valence and sea masses.  We
 can easily generalize from there to arbitrary $N$ with arbitrary sea-quark  masses.   We will not attempt to
 prove the result for arbitrary $N_v>2$, but will argue that it is probably true in that case too.
 
 In the neutral sector of the $N=2$, $N_v=1$  model, there are 3 mesons, $\pi$, $\delta$, and $\epsilon$,
 with the meson field $\Phi$ of \eq{PQSigma} given by
 \be\eqn{Phi21}
 \Phi = \frac{1}{\sqrt{6}}\diag\left(\sqrt{3}\pi+ \delta  +i \epsilon  ,\;- \sqrt{3}\pi+ \delta  +i \epsilon  , \;-2 \delta  +i \epsilon  ,\; 3i \epsilon  \right),
 \ee
 where entries are ordered sea, valence, ghost.
  The quark mass matrix is $\cM=\diag(m,m,m_x,m_x)$.   Expanding the Lagrangian in momentum space to quadratic order in these
  fields, we find 
  \be\eqn{L21-quad}
  \cL_{\rm quad} = \frac{1}{2} (p^2+M_\pi^2)\, \pi^2 + \frac{1}{2} (p^2+M_\delta^2)\, \delta^2+ \frac{1}{2} (p^2+M_\epsilon^2)\, \epsilon^2 + i\alpha \delta\epsilon,
  \ee
  where
  \bea
  M_\pi^2 &=& B_0(2m), \hspace{2.4cm}   M_\delta^2 = B_0\left(\frac{4}{3}m_x +\frac{2}{3}m \right), \nonumber \\
 M_\epsilon^2 &=& B_0\left(\frac{8}{3}m_x -\frac{2}{3}m \right), \qquad  \alpha = B_0\left(\frac{2}{3}m -\frac{2}{3}m_x\right).
 \eea
 We see that $M^2_\epsilon$ is only positive for $m_x > m/4$, which is precisely the Sharpe-Shoresh lower
 bound for this case. However, there is also an  imaginary (hence not Hermitian) mixing term between $\epsilon$
 and $\delta$. (There is no mixing with $\pi$ in this model because of the exact sea-quark isospin symmetry.)
 
 The $ \epsilon$-$\delta$ mixing term has an important effect:  If we treat it as an iterated 2-point interaction, it generates the expected double poles for neutral particles in a partially-quenched theory.  In other words,
 it plays the role that the anomalous $\Phi_0$ mass term plays in 
 the case where the limit of infinite $\Phi_0$ mass
  is postponed until after the computation of the neutral propagators. Note that the poles (whether single or double)
of a partially-quenched neutral propagator occur at the squared-masses  either of physical sea-sea neutral mesons, or of  unmixed valence-valence meson masses, which here would be proportional to $m_x$.  All
these squared-masses are positive (for nonzero quark masses), which suggests that the apparent problem
of $M^2_\epsilon <0$ can be avoided if we treat the mixing term $i\alpha\delta\epsilon$ on the same footing
as the other mass terms.  We can accomplish this by performing the path integral over $\delta$ {\em before}\/ the 
$\epsilon$ integral.  The $\delta$ integral  is convergent
since $M^2_\delta>0$.

By completing the square, we may integrate over $\delta$ in the partition function
\be\eqn{Zorig}
Z = \int D\pi\; D\epsilon\; D\delta\; e^{-\int d^4x \, \cL_{\rm quad}}
\ee
and obtain
\be\eqn{Zeta-out}
Z = \int D\pi\; D\epsilon\;  e^{-\frac{1}{2}\int d^4x \left( (p^2+M^2_\pi)\pi^2 + F(p,M_\epsilon,M_\delta,\alpha)\epsilon^2\right)},  
\ee
 where
 \bea\eqn{F}
 F(p,m_\epsilon,m_\delta,\alpha) &=& p^2 + M_\epsilon^2+\frac{\alpha^2}{p^2+M_\delta^2}= \frac{(p^2+M_X^2)^2}
 {p^2+M^2_\delta}, \\
 \eqn{MX}
 M_X^2 &\equiv& B_0(2m_x).
 \eea
 Here $M_X$ is the mass of the valence 
 $\bar x x$ meson.  Note that $F$ is positive definite, so the $\epsilon$ integral
 converges, independent of whether $m_x$ is above or below the Sharpe-Shoresh lower bound. 
 
 The $\epsilon$
 propagator, $G_{\epsilon\epsilon} = F^{-1}$ has a characteristic double pole at $M^2_X$:
 \be\eqn{eps-prop-21}
 G_{\epsilon\epsilon} = \frac{p^2+M^2_\delta}{(p^2+M_X^2)^2} = \frac{1}{3}\left[\frac{2}{p^2+M_X^2}
 +\frac{p^2+M^2_\pi}{(p^2+M_X^2)^2}\right],
 \ee  
 where the second form will simplify the comparison to more complicated cases.
 We can also obtain
 this result  for $G_{\epsilon\epsilon} $  by iterating the $i\alpha\delta\epsilon$ term in \eq{L21-quad} or
 by returning to the theory with $\Phi_0$ included, iterating the $\Phi_0$ mass term, and then taking the mass
 to infinity.  Repeating the derivation of \eq{Zeta-out} with sources for the $\delta$ and $\epsilon$ fields
 allows us to find the $\delta$ propagator $G_{\delta\delta} $  and the mixed propagator $G_{\epsilon\delta} $, which
 of course also agree with the results found using the other methods. 
 
 We have thus obtained a convergent $\epsilon$ path integral by integrating over
 $\delta$  first.  The path
 integral, as given by  \eq{Zorig},  is however only conditionally convergent for  $m_x<m/4$. 
  The integral of
 the absolute value is not convergent, since the imaginary mixing term is lost and the $\epsilon$ integral
 is then divergent.  Alternatively, it is clear that
 the result is ill defined for  $m_x<m/4$ if the
  $\epsilon$ integral in \eq{Zorig} is performed first.

  The lack of absolute 
  convergence of \eq{Zorig} suggests that the partially quenched chiral 
  theory for $m_x<m/4$ is a delicate object for which the usual manipulations of perturbation theory are
  suspect and must be checked carefully.  
  This suggestion is however incorrect!  The reason is
  that  the exponent in \eq{Zorig} is only the quadratic action, and not the full action.  With the full
  action, the $\epsilon$ integral is convergent even if it is performed first.  That is because the
  ghost-ghost component of $\Phi$ in \eq{Phi21} gives rise to a term in the potential that
  goes like $+m_x\cosh(\sqrt{6}\,\epsilon/f)$, which dominates for large $|\epsilon/f|$ over the negative 
  term $-(2m+m_x)\cosh(\sqrt{2/3}\,\epsilon/f)$ for all nonzero 
  values of $m_x$ and $m$.  Thus if we make the quadratic
  approximation of \eqs{L21-quad}{Zorig}, we should consider the $\epsilon$ integral as cut off at large
  but finite $|\epsilon|$.  (We might, for example, add a term $\lambda \epsilon^4$ to the Lagrangian,
  where $\lambda$ is small and positive.) In that case the full path integral is absolutely convergent, and by
  Fubini's theorem \cite{Fubini} the order of
  integration does not affect the answer.   It is true, though, that the simplest way to evaluate the partition function
   is by  first integrating  over $\delta$. Indeed,  we do not know how
  to show that the final result even has a limit when $\lambda\to0$, except by the use of Fubini's theorem.

 In the
 presence of nonzero, but small, $\theta$, these results change only slightly.  The saddle point
 is now at $\pi=0$, $\delta=\delta_0$, $\epsilon=i\delta_0$, where $\delta_0$ is of order $\theta$.  For
 $m^2_\epsilon>0$ the steepest descent directions from the saddle point are the real directions for
 all three fields, and we keep those directions even when $m^2_\epsilon<0$.  Expanding around the saddle,
 there are of course no linear terms in the fields,  and the quadratic terms are changed slightly by
 cosines of the angles $\theta$ and $\delta_0$.  The end result is that \eqs{Zeta-out}{F}
 are still valid, but with redefined values:
 \bea
 \hspace{-2mm}M_X^2 & =& 2B_0\sqrt{m_x^2-m^2\sin^2(\theta/2)}, \nonumber \\
   M^2_\delta &=& B_0\!\left(\frac{2}{3} m\cos(\theta/2) +\frac{4}{3} 
 \sqrt{m_x^2-m^2\sin^2(\theta/2)}\right).\eqn{masses-theta}
 \eea
We may restrict  ourselves to  infinitesimal $\theta$ values in order to find the
 derivatives with respect to $\theta$ at $\theta=0$. The quadratic action
 of $\epsilon$ therefore remains positive definite for any nonzero quark masses.  
 On the other hand, for $\theta$ small
 but finite, a new singularity would develop for  very small but nonzero $m_x < m\sin(\theta/2)$.  
 We do not concern ourselves further with this interesting, but to us irrelevant,  singularity.
 
 We now turn to the case discussed in \secref{PQ}:  $N=3$ (masses $m_u=m_d=m$, $m_s$)
 and $N_v=2$ (masses $m_x$, $m_y$).  In this case, there are two neutral ghost-type fields, corresponding 
 to $\epsilon$ and $\gamma$ in \eq{vacuum-general}, and three 
 neutral  quark-type fields that they may mix with, corresponding to $\alpha$, $\beta$, and $\delta$.  Because
 the algebra involved in doing the  convergent integrals over $\alpha$, $\beta$, and $\delta$ is rather messy,
 we resort to a short cut.  As noted  in the discussion of the toy model above,  Gaussian integration (with possible linear terms) is equivalent to
 perturbation theory. Hence we can simply find the $\epsilon$-$\epsilon$, $\gamma$-$\gamma$ and 
 $\epsilon$-$\gamma$ propagators using perturbation theory, and deduce the positivity of the 
 $\epsilon$-$\gamma$ quadratic action from the propagator matrix.   
 
 The easiest way to obtain the needed propagators is to restore the anomalous $\Phi_0$ field (mass $m_0$)
and use the diagonal basis for neutral fields, $\Phi = \diag(U,D,S,X,Y,\tilde X, \tilde Y)$.  Quark-line connected and 
disconnected propagators among the diagonal-basis fields are the standard ones of PQ\chpt, and the
limit $m_0\to\infty$ may be taken.  We then just need to write $\epsilon$ and $\gamma$ as linear combinations of 
the diagonal-basis fields, and take the corresponding linear combinations of the propagators.
With
\bea\eqn{epsilon}
\epsilon &=& -i\sqrt{\frac{2}{15}}\left(U+D+S+X+Y+\frac{5}{2}\tilde X+ \frac{5}{2}\tilde Y\right), \\
\gamma &=& -i\sqrt{\frac{1}{2}}\left(-\tilde X +\tilde Y\right) \eqn{gamma},\\
M_X^2 &=& 2B_0m_x,\quad M_Y^2 = 2B_0m_y,\quad M_\pi^2 = 2B_0 m,\nonumber\\
M_S^2 &=& 2B_0m_s,\quad M_\eta^2 = B_0(2m/3 + 4m_s/3),\eqn{masses}
\eea
we find
\bea
\hspace{-3mm}G_{\epsilon\epsilon} &=& \frac{1}{10}\left[  \frac{3}{p^2+M^2_X} + \frac{3}{p^2+M^2_Y} +
\frac{(p^2+M_\pi^2)(p^2+M_S^2)}{p^2+M^2_\eta}\left(\frac{1}{p^2+M^2_X} + \frac{1}{p^2+M^2_Y}\right)^2\right],  
\nonumber\\
\eqn{G}
\hspace{-3mm}G_{\gamma\gamma} &=& \frac{1}{6}\left[  \frac{3}{p^2+M^2_X} + \frac{3}{p^2+M^2_Y} +
\frac{(p^2+M_\pi^2)(p^2+M_S^2)}{p^2+M^2_\eta}\left(\frac{1}{p^2+M^2_X} -\frac{1}{p^2+M^2_Y}\right)^2\right],  
\\
\hspace{-3mm}G_{\epsilon\gamma} &=& \frac{1}{\sqrt{60}}\left[  \frac{3}{p^2+M^2_Y} - \frac{3}{p^2+M^2_X} +
\frac{(p^2+M_\pi^2)(p^2+M_S^2)}{p^2+M^2_\eta}\left(\frac{1}{(p^2+M^2_Y)^2} -\frac{1}{(p^2+M^2_X)^2}\right)\right].
\nonumber
\eea

We write the  $\epsilon$-$\gamma$ propagator matrix as
\be\eqn{Gmatrix}
G = \begin{pmatrix}
G_{\epsilon\epsilon} & G_{\epsilon\gamma} \\
G_{\epsilon\gamma} & G_{\gamma\gamma}
\end{pmatrix}
\ee
Then the $\epsilon$-$\gamma$ quadratic action matrix, after integrating out all the neutral quark-antiquark fields
that have quadratic interactions with $\epsilon$,\footnote{No such fields 
have quadratic interactions with $\gamma$} must be $G^{-1}$. 
Because $G$ and $G^{-1}$ are real symmetric matrices, they are diagonalizable.  This means that if one
of them is positive definite, the other must be too.   So we need only prove that $G$ is positive definite.
We can do that by showing that its eigenvalues are both positive, which simply requires $\tr(G)>0$ and
$\det(G)>0$.   From \eq{G}, $\tr(G)$ is clearly positive.
After some algebra, we find
\be\eqn{det}
 \hspace{-3mm}\det(G) = 
   \frac{1}{5(p^2+M^2_X)(p^2+M^2_Y)}\left[3+
\frac{(p^2+M_\pi^2)(p^2+M_S^2)}{p^2+M^2_\eta}\left(\frac{1}{p^2+M^2_X} +\frac{1}{p^2+M^2_Y}\right)\right],
 \ee
 which is also positive.  
 
 Thus,  the quadratic action of the neutral ghost-antighost fields in the $N=3$, $N_v=2$
 case is positive definite after integration of the neutral quark-antiquark fields at quadratic order.
 As in the $N=2$, $N_v=1$ case, it also true that here that the quadratic ghost-antighost integrals may
 be considered to be cut off at large field values by higher terms in the full action.  The cutoff
terms, which come from contributions to the potential from the ghost-ghost block, grow like $m_x\cosh\left((\sqrt{10/3}\,\epsilon-\sqrt{2}\,\gamma)/f\right) +
  m_y\cosh\left((\sqrt{10/3}\,\epsilon+\sqrt{2}\,\gamma)/f\right)$ and dominate the negative quark-quark contributions
  in all real directions of the $\epsilon$-$\gamma$ plane.  So, once again, the path integral around the
  trivial vacuum to quadratic order is absolutely convergent.  The order of integration has no effect,
  except in the practical sense that integrating the neutral quark-antiquark fields first is much easier.
 
 For  small but nonzero $\theta$ the analysis follows the procedure described above for 
 $N=2$, $N_v=1$.  The saddle point occurs
 at imaginary values of the neutral ghost-antighost fields.  We expand around the saddle point in
 the real directions of these fields.  Compared to the $\theta=0$ case, the quadratic terms are changed
 slightly by cosines of linear combinations of $\theta$ and the saddle-point angles,  
 which can make small differences in the meson masses, 
 as in \eq{masses-theta}.  The theory therefore remains positive except possibly for very small
 values of the valence masses of order $\bar m \theta$  
(where $\bar m$ is the average sea-quark mass).
Such possibilities do not pose any difficulties for our analysis in the body of this paper, since we only need to consider infinitesimal $\theta$ values 
 to find the derivatives at $\theta=0$.

 It is  quite easy to generalize our results to an arbitrary number $N\ge2$ of sea-quark flavors, 
 and arbitrary sea-quark masses. The only changes in \eqs{G}{det} will be (1) an adjustment of the relative normalization of the single
 pole and double pole terms) and (2) a replacement of the factor 
 $(p^2+M_\pi^2)(p^2+M_S^2)/p^2+M^2_\eta)$ in each equation  with  the corresponding sea-meson factor that multiplies the disconnected neutral propagator in the given theory.  Since
 this factor ``goes along for the ride'' in all the manipulations that led to \eqs{G}{det},  the quadratic action in the ghost-ghost sector will remain positive definite.
 
 Since $N_v=2$ is the most useful case for analyzing simulations, we have not tried very hard to generalize to
 the more complicated cases with $N_v>2$. 
 There are however some indications that the quadratic action of the ghost-antighost 
 sector remains positive
 definite after integration of the neutral quark-antiquark fields. 
  First of all, it is clear that $G_{\epsilon\epsilon} > 0$, since the sum of two valence-mass poles
 in \eq{G} will just be replaced by the sum of $N_v$ valence mass poles (and again the relative normalization
 of single and double pole terms may change).  We can see this change explicitly by comparing with the $N_v=1$ 
 case, \eq{eps-prop-21}.  
 
 Second, when the valence masses are degenerate, $\epsilon$ does not interact
 in the quadratic action with any other ghost-type fields (which themselves have positive
 masses and do not interact with the neutral quark-type fields). Therefore the positivity of $G_{\epsilon\epsilon}$
 is all we need for positivity of $G$.  Any nonpositivity in the nondegenerate case would have to come from large effects of interactions with the other ghost-type fields.  Such effects can occur before integration over the quark-type fields, as in the Sharpe-Shoresh bound. In that case $M^2_\epsilon$ can be positive but small, and then interactions can generate a negative eigenvalue in the complete ghost-type mass matrix.   However, this seems  less likely to occur after integration over the quark-type fields.  Based on our $N_v=2$ example
 in \eq{G}, it looks
 difficult to choose masses such that $G_{\epsilon\epsilon}$ is small, but interaction
 terms (such as $G_{\epsilon\gamma}$) get  large enough to change the sign of an eigenvalue.  
In any case it is clear that positivity will be guaranteed in some neighborhood of the degenerate point.

\section{Decoupling}
\label{decoupling}

If we take one quark mass to infinity and look at a meson made of light quarks,
the decoupling paradigm implies that the meson should be unaffected by the heavy quark.  The results
in \secrefs{nf2}{nf3} give an example of this.
Taking $m_s\to\infty$ in the pion mass of \eq{mpi-nf3}, gives
$M_\pi'' = -M_\pi(0)/8$.  This agrees with the result for the degenerate
($m_u=m_d$) 
pion in $N=2$, \eq{mpi2pp}.  The agreement is nontrivial, since the light-light part of the
chiral Lagrangian starts with different $\theta$-dependence in the two cases:  $\exp(i\theta/3)$
versus  $\exp(i\theta/2)$.   The decay constant results trivially
agree with decoupling, since they vanish in both cases for degenerate light
quarks.

For a more robust check of decoupling, we look at the
case of arbitrary $N\ge3$ quarks,  with $k<N$ light degenerate quarks
of mass $m$ and $N-k$ degenerate heavy quarks of mass $m_h$.
In the first instance, we proceed exactly as in \secrefs{nf2}{nf3}, starting with a factor of $e^{i\theta/N}$
multiplying all quark masses, both light and heavy, as in \eq{ChPT-L}.
Let $\langle\Sigma\rangle$ be given by a generalization of \eq{Sigma-nf3},
with $\exp(i\alpha)$ in the first $k$ diagonal entries, and
$\exp(-ik\alpha/(N-k))$ in the last $N-k$ diagonal entries.  
Following the same steps as before, we obtain
\begin{equation}
\eqn{alphap-arb-nf}
\alpha'= \frac{m-m_h}{Nm+Nkm_h/(N-k)}.
\end{equation}
As always, the decay constant for the degenerate quarks vanishes, while
the mass obeys
\begin{equation}
M_\pi'' =
-\frac{M_\pi(0)}{2}\;\frac{m^2_h}{\big((N-k)m+km_h\big)^2}.
\eqn{mpi-arb-nf}
\end{equation}

Alternatively, following the discussion in Appendix \ref{m=0}, we may make a non-anomalous transformation
at the start, so that  we have a factor of $e^{i\theta/k}$ multiplying the light quark masses only. The vacuum
expectation value $\langle\Sigma\rangle$ has the same form in terms of $\alpha$ as above, but now
\begin{equation}
\eqn{alphap-arb-nf-alt}
\alpha'= \frac{m}{km+k^2m_h/(N-k)}.
\end{equation}
As expected, however, \eq{mpi-arb-nf} is unchanged.

In either case, taking $m_h\to\infty$ gives $M_\pi''=-M_\pi(0)/(2k^2)$, independent
of the existence or number of heavy quarks.  The same answer would be obtained
if we had set $k=N$ from the beginning, \ie if we had only (degenerate) light quarks.%
\footnote{In that case we must immediately
put $\alpha=0$ because nonzero $\alpha$ is inconsistent with the dual requirements of flavor symmetry
and $\det(\Sigma)=1$.}   Thus, no matter how the problem is set up, the heavy quarks decouple when
$m_h\to\infty$. The result in that limit is the same as it would be if the heavy quarks were
not there at all. Note, however, that the decoupling is subtle:  solving the problem with the heavy
quarks present and then taking their mass to infinity is not a matter of simply deleting all terms in the 
original Lagrangian that involve the heavy quarks.  This is particularly clear when we set up the problem
with $e^{i\theta/N}$
multiplying all quark masses.  Simply deleting the heavy quark terms after the
set-up would give the incorrect result $M_\pi''=-M_\pi(0)/(2N^2)$.

\end{document}